\documentclass[10pt,onecolumn]{IEEEtran}

\usepackage{framed}
\usepackage{color}
\usepackage{multirow}

\definecolor{bgrd}{rgb}{1,1,1}
\definecolor{grey}{rgb}{0.9,0.9,0.6}
\definecolor{gray}{rgb}{0.5,0.5,0.5}
\definecolor{dkr}{rgb}{0.6,0.2,0.2}
\definecolor{dkg}{rgb}{0,0.5,0}
\definecolor{dkb}{rgb}{0.0,0.1,0.7}
\definecolor{light-gray}{gray}{0.85}

\renewcommand{\P}{\mathbb{P}}

\newcommand{\beq}{\begin{equation}}
\newcommand{\eeq}{\end{equation}}
\newcommand{\beqa}{\begin{eqnarray}}
\newcommand{\eeqa}{\end{eqnarray}}
\newcommand{\dfz}{\stackrel{\Delta}{=}}

\newcommand{\bX}{{\mathbf{X}}}

\newcommand{\bx}{{\mathbf{x}}}

\newcommand{\cM}{{\cal M}}

\newcommand{\E}{\mathbb{E}}

\newcommand{\cH}{{\cal H}}

\newcommand{\cP}{{\cal P}}
\newcommand{\cQ}{{\cal Q}}

\newcommand{\cX}{{\cal X}}

\newcommand{\cI}{{\mathbb I}}
\usepackage{cite}
\usepackage{amssymb}
\usepackage{amsbsy}
\usepackage{mathrsfs}
\usepackage{graphicx}
\usepackage{epstopdf}
\usepackage{amsmath}

\usepackage{subfigure}

\usepackage{soul}
\usepackage[linesnumbered,ruled,vlined]{algorithm2e}

\begin{document}


\title{Algorithms and Fundamental Limits for \\ Unlabeled Detection using Types}

\author{Stefano~Marano, and Peter Willett, \emph{Fellow, IEEE}
\thanks{S.~Marano is with DIEM, University of Salerno, via Giovanni Paolo~II 132, I-84084, Fisciano (SA), Italy (e-mail: marano@unisa.it)
P.~Willett is with ECE Dept., University of Connecticut, Storrs, CT, USA (e-mail: peter.willett@uconn.edu).}
\thanks{P.\ Willett was supported by NPS via ONR contract N00244-16-1-0017.}
\thanks{Part of this work has been presented at the 2018 IEEE International Conference on Acoustics, Speech and Signal Processing (ICASSP 2018)
15-20 April 2018, Calgary, Alberta, Canada~\cite{ICASSP2018unlab}. This paper offers an improved expression that should be considered a replacement for that in~\cite{ICASSP2018unlab}.}
}
\maketitle

\begin{abstract}

Emerging applications of sensor networks for detection sometimes suggest that classical problems ought be revisited under new assumptions. 
This is the case of binary hypothesis testing with independent 
-- but not necessarily identically distributed -- observations under the two hypotheses, 
a formalism so orthodox that it is used as an opening example in many detection classes.
However, let us insert a new element, and address an issue perhaps with impact on strategies to deal with ``big data'' applications: What would happen if the structure were streamlined such that data flowed freely throughout the system without provenance? How much information (for detection) is contained in the sample \emph{values}, and how much in their \emph{labels}? How should decision-making proceed in this case? 
The theoretical contribution of this work is to answer these questions by establishing the fundamental limits, in terms of error exponents, of the aforementioned binary hypothesis test with \emph{unlabeled} observations drawn from a finite alphabet. Then, we focus on practical algorithms. 
A low-complexity detector --- called ULR ---  solves the detection problem without attempting to estimate the labels. A modified version of the auction algorithm is then considered, and two new greedy algorithms with ${\cal O}(n^2)$ worst-case complexity are presented, where $n$ is the number of observations. The detection operational characteristics of these detectors are investigated by computer experiments.

\end{abstract}

\begin{IEEEkeywords} 
Unlabeled detection, fundamental limits of hypothesis testing, error exponents, types, assignment problem, greedy algorithms.
\end{IEEEkeywords}

\section{Introduction and Motivations}
\label{sec:intro}

Mostly motivated by emerging applications of sensor networks, recent years have seen the birth of a field that can be referred to as \emph{signal processing with unlabeled data}. This terminology refers to the bulk of classical algorithms and methods of signal processing, revisited under the new paradigm of a central unit that must process a vector of data received by certain peripheral units, but must do so -- or choose do so --  without access to the data \emph{labels}, namely without knowing the original position of each datum inside the vector. The meaning here given to ``labeling'' is that of \emph{provenance} and is not to be confused with the labeling obtained by data classification, as typical, for instance, of machine learning applications. Note that in this work we are interested in the first case, that the processing must proceed without labels by necessity; when labeling is avoided as a matter of elegance is usually referred to as the random finite set (RFS) idea, and good entry points are \cite{mahler_book,mahler_tutorial}.

As a notional example, suppose that under the null hypothesis two sensors' observations are independent and identically distributed (\emph{iid}) 
unit-normal; and that under the alternative their means are shifted, respectively by $+1.2$ \& $-1.2$. The central decision-maker receives the set $\{-1.3,+1.3\}$, and is specifically told that it should make no assumption about which observation came from which sensor. Intuition suggests that the first sensor's observation is $+1.3$ and the second sensor saw $-1.3$; and hence that there is a fairly decent fit with the alternative hypothesis. 
How much decision-making
performance has been lost by  {\em label-agnostic} decision-making with respect to  {\em label-aware} in this case? That is, how much information is in knowing who said what, as opposed simply to knowing what was said?
And how about the case that the two mean shifts were respectively $1.1$ \& $1.3$: clearly the quality of the match is much lower; but equally clearly the impact of making a labeling error is far lower.

\subsection{Related Work}

Modern networks are vulnerable to malicious attacks. For instance, the civilian global positioning system (GPS) is particularly exposed to spoofing attacks~\cite{Humphreys-2008}, which can impair wireless ad-hoc communication systems~\cite{6415619-Milcom2012}, or alter the timing information in smart grids~\cite{7860525CNS2016,6400273-SMARTGRID2013}. 
As a consequence, the timestamp information of the system may be altered to the point that the data arriving to a central decision unit can be considered unlabeled.

Even in absence of an attack, modern sensor networks and other networked inference/communication systems are similarly vulnerable, especially when faced with big-data applications. Indeed, one challenge of these systems is the possible presence at the fusion center of data partially unordered --- the so-called out-of-sequence measurements (OOSM) issue.
A prominent example is represented by distributed tracking systems where the data received at the fusion center are partially unordered~\cite{Challa2003}. Similarly, networked control systems with packetized messages can be subject to various timing errors due to uncontrollable packet delays~\cite{4608943-AC2008}. In~\cite{7266584-BracaWillett2015} the lack of a precise timestamp of data is considered in connection with the usage of automatic identification system (AIS) in real-world maritime surveillance problems.
The common denominator to all these examples is that data must be processed with partial or no information about their relative time/space ordering, which is ofter related to their provenance from a peripheral unit of the network.

A systematic study of the lack-of-provenance issue, which is nowadays referred to as the \emph{unlabeled data} paradigm, has been prompted by~\cite{Vetterli2018IT,7447086-ALLERTON2015}. The authors of~\cite{Vetterli2018IT,7447086-ALLERTON2015} consider a signal recovery problem from a set of unlabeled linear projections. They also compare their unlabeled sensing formulation with the setting of compressed sensing (see e.g.,~\cite{eldar,6853755-ICASSP2014}), and highlights connections with a classical problem in robotics which is known as simultaneous location and mapping (SLAM)~\cite{SLAM}.
Very recent studies with a similar data-reconstruction focus can be found in~\cite{8234626-Caire2018SP,Abid2017,Pananjady2017,7852261-ALLERTON2016,Pananjady2016}.

In contrast to data reconstruction, our focus is on \emph{inference} by unlabeled data, which has been addressed in the last few years by~\cite{WangSP2018,7904659-CommLett2017,marano-ICASSP17}. 
In particular, we elaborate on a model similar to that addressed in~\cite{marano-ICASSP17}, under the assumption that data are drawn from a finite alphabet.
The motivation is that modern applications of large wireless sensor networks frequently impose severe constraints on the delivered messages, due to limited sensors' resources, e.g., energy, bandwidth, etc. 
In these applications, to include the identities of the reporting sensors in the delivered messages might constitute an excessive burden~\cite{keller-2009} and, for the same reasons, the delivered data are usually constrained to belong to a finite alphabet with small cardinality.

\vspace*{-3pt}
\subsection{Contribution}

To illustrate our contribution, consider the already mentioned works~\cite{Vetterli2018IT,7447086-ALLERTON2015}. There, the authors find a fundamental limit for data reconstruction: if only \emph{unlabeled} linear projections are observed, a perfect data recovery of a $n$-vector $\bx^n=(x_1,\cdots,x_n)$ is possible provided that the number of such projections is at least $2n$. Conversely, if this number is less than $2n$, there is no way to recover the original $\bx^n$ from its projections. Doubling the size is the fundamental limit for data reconstruction.
Note, in passing, that the factor 2 is reminiscent of a fundamental result in compressed sensing theory, see~\cite{ludo,eldar}.
One goal of this paper is to develop a similar fundamental limit for binary detection, instead of reconstruction, from unlabeled data. 
To be concrete, suppose that the divergence between the data distributions under the two hypotheses is taken as a proxy of the asymptotic ($n\rightarrow \infty$) theoretical optimal detection performance 
when one observes the vector $\bx^n$.
We pose the question: what is the optimal theoretical detection performance in situations where only an unordered version of $\bx^n$ is observed, namely, when 
we know the values of the entries of $\bx^n$ but not their ordering? 
How much information for detection is contained in the entry \emph{labels} and hence is lost, and how much in the entry \emph{values}, and hence retained by the unlabeled version of $\bx^n$?
The notional example presented above suggests that even the unlabeled version of $\bx^n$ carries some information for detection, 
but no much more than this na\"{i}ve notion is known. We fill this gap for a class of detection problems that will be formalized in~(\ref{eq:testun}).

After answering these questions we make a step further. Characterizing the ultimate detection performance does not tell very much about the possibility of 
solving the unlabeled detection problem with \emph{practical} detectors. This motivates us to investigate if there exist detection algorithms with affordable computational complexity and acceptable performance 
for finite values of $n$. First, we show that the unlabeled detection problem with discrete data can be recast in the form of a classical assignment problem, for which optimal algorithms are known, but can be highly inefficient for our problem.
Then, we develop two new algorithms which require lower computational complexity. Computer simulations are presented to assess the detection performance and the computational burden of these detectors.

The remainder of this paper is organized as follows. The next section introduces the classical setup of detection with labeled data. Section~\ref{sec:unlabeled} formalizes the unlabeled detection problem and presents the main theoretical results. Practical algorithms for unlabeled detection are considered in Sec.~\ref{sec:twodet}, while the results of computer experiments are 
presented in Sec.~\ref{sec:compe}. Section~\ref{sec:conc} concludes the paper. Some technical material is postponed to Appendices~\ref{app:P1}-\ref{app:wp1}.

\section{Classical Detection with Labeled Data}
\label{sec:lab}

Let $\bX^n=(X_1,\dots,X_n)$ be a vector whose entries are~$n$ random variables defined over a common finite alphabet~$\cX$, and let $\bx^n=(x_1,\cdots,x_n)$ be the correspondent realization. We focus on the asymptotic scenario of $n\rightarrow \infty$, and is therefore appropriate to add a superscript~$^n$ to specify the size of the vectors. 
Also, let $\cP(\cX)$ denote the set of all probability mass functions (PMFs) on $\cX$. 
As usual, $\cX^n$ denotes the $n$-th extension of the alphabet $\cX$, namely, the concatenation of $n$ letters from $\cX$, and  $\cP(\cX^n)$
denote the set of PMFs over $\cX^n$. 

The binary hypothesis test we consider is as follows. 
Under hypothesis $\cH_0$ the joint probability $q_{1:n}(\bx^n)$ of vector $\bX^n$ is the product of possibly non-identical marginal PMFs $q_{1:n}(\bx^n) = \prod_{i=1}^n q_i(x_i)$, where $q_i \in \cP(\cX)$. Likewise,
under $\cH_1$ the joint probability $p_{1:n}(\bx^n)$  is the product of possibly non-identical marginal PMFs $p_{1:n}(\bx^n) = \prod_{i=1}^n p_i(x_i)$, with $p_i\in \cP(\cX)$. This means that  data are independent but not necessarily identically distributed under both hypotheses. 
Formally, we have 
\beq
\bX^n \sim r_{1:n}(\bx^n) = \prod_{i=1}^n r_i(x_i) \;\;\; \left \{
\begin{array}{l}
\cH_1 :  r_i(x_i)= p_i(x_i), \\ 
\cH_0 :  r_i(x_i)= q_i(x_i),
\end{array}
\right .
\label{eq:test}
\eeq
for $n=1,2,\dots$.
It is assumed throughout that $q_i(x)>0$ and $p_i(x)>0$, for all $i=1,2,\dots,n$, and all $x \in \cX$. This simplifies some results and excludes the singular cases in which the test can be solved without error for $n\rightarrow \infty$.

The Kullback-Leibler divergence from $q_i(x)$ to $p_i(x)$ is defined as~\cite{CT2}: $D(q_i \| p_i) \dfz \sum_{x\in\cX} q_i(x) \log \frac{q_i(x)}{p_i(x)}$,
and the assumption of strictly positive PMFs implies that~$D(q_i \| p_i)$ exists and is finite for all~$i$. All logarithms are to base $e$.

The error probabilities of  test~(\ref{eq:test}) are
\beqa
\P_0( \bX^n \not \in A_n)&&  \textnormal{type I error}, \\
\P_1( \bX^n \in A_n)&& \textnormal{type II error},
\eeqa
where $A_n \subseteq \cX^n$ is some decision region in favor of $\cH_0$, and $\P_h$ is the probability operator under $\cH_h$, $h=0,1$.

For two sequences of distributions\footnote{We often simplify the notation by omitting the argument $\bx^n$: we simply write $q_{1:n}$, $p_{1:n}$, for $q_{1:n}(\bx^n)$, $p_{1:n}(\bx^n)$, and similar.} $q_{1:\infty},p_{1:\infty}\in \cP(\cX^n)$, let us define the
\emph{divergence rate} 
\beq
\bar{D}(q_{1:\infty}\|p_{1:\infty}) \dfz \lim_{n\rightarrow \infty} \frac 1 n \sum_{i=1}^n D(q_i \| p_i).
\eeq
We assume that the divergence rates encountered in this paper exist, are finite, and are continuous and convex functions of their arguments. 
This is a very mild requirement that rules out pathological 
choices of the sequences $p_{1:\infty}, q_{1:\infty}$, which are of no practical interest. 
Let us introduce now the error exponent function, and then state two classical results about the asymptotic error exponents of the hypothesis test.

\vspace*{3pt}
\noindent
\textsc{Definition} 
\emph{\emph{(Error Exponent for Labeled Data)}:
For $\alpha>0$, let us define
\beq
\Omega_{\rm lab}(\alpha) \dfz \hspace*{-10pt} \inf_{\omega_{1:\infty} \in \cP(\cX^\infty): \, \bar{D}(\omega_{1:\infty}\|q_{1:\infty}) < \alpha} 
\hspace*{-10pt} \bar{D}(\omega_{1:\infty} \| p_{1:\infty}).
\label{eq:Psi}
\eeq\hfill} 

\noindent It is useful to bear in mind that $\Omega_{\rm lab}(\alpha)$ depends on the sequences $q_{1:\infty}$ and $p_{1:\infty}$. When needed, we use the more precise notation $\Omega_{\rm lab}(\alpha; p_{1:\infty}, q_{1:\infty})$.

\vspace*{3pt}
\noindent
\textsc{Proposition 1} 
\emph{\emph{(Labeled Detection~\cite{blahut-HT&IT})}
Consider the hypothesis test~(\ref{eq:test}). 
Let $0 < \alpha < \infty$.
\begin{itemize}
\item[a)] Let $A_n \subseteq \cX^n$ be any sequence of acceptance regions for~$\cH_0$. Then:
\begin{align}
& \liminf_{n\rightarrow \infty} - \frac 1 n \log \P_0(\bX^n \not \in A_n) \ge \alpha \nonumber \\
& \Rightarrow \; \limsup_{n\rightarrow \infty} - \frac 1 n \log  \P_1(\bX^n  \in A_n) \le \Omega_{\rm lab}(\alpha).
\label{eq:class1}
\end{align}
\item[b)] There exists a sequence $A_n^\ast\subseteq \cX^n$ of acceptance regions for~$\cH_0$ such that
\begin{subequations}\begin{align}
&\liminf_{n\rightarrow \infty} - \frac 1 n \log \P_0(\bX^n \not \in A_n^\ast)\ge \alpha, \label{eq:class3} \\
&\lim_{n\rightarrow \infty} - \frac 1 n \log  \P_1(\bX^n \in A_n^\ast) = \Omega_{\rm lab}(\alpha). \label{eq:class4}
\end{align}\end{subequations}
\end{itemize}}

\vspace*{2pt}\noindent\emph{Proof:} This is a standard result and the proof is sketched in Appendix~\ref{app:P1}\hfill~$\bullet$\vspace*{3pt}

Part $a)$ of the proposition states that whatever the sequence of decision regions is, if type~I error goes exponentially to zero at rate not smaller than~$\alpha$, then type~II error goes to zero exponentially at rate not larger than $\Omega_{\rm lab}(\alpha)$.
Part $b)$ states that the above limits are tight in the sense that there exists a sequence $A_n^\ast$ of decision regions such that the best rate $\Omega_{\rm lab}(\alpha)$ for type~II error is achieved.

For two sequences $a_n$ and $b_n$, the symbol
$a_n \stackrel{\cdot}{=} b_n$ means equality to the first order in the exponent, namely $\lim_{n\rightarrow \infty} \frac 1 n \log \frac{a_n}{b_n}=0$.
We can summarize the content of Proposition~1 by saying that for problem~(\ref{eq:test}) it is possible to find tests such that type~I error is 
$\stackrel{\cdot}{=} e^{-n \alpha}$ and type~II error is $\stackrel{\cdot}{=} e^{-n \Omega_{\rm lab}(\alpha)}$,
but no stronger pairs of asymptotic expressions can be simultaneously verified. 
Note that $\lim_{\alpha \rightarrow 0} \Omega_{\rm lab}(\alpha)=\bar{D}(q_{1:\infty} \| p_{1:\infty})$. The following standard result
emphasizes the operational meaning of this divergence rate.

\vspace*{3pt}
\noindent
\textsc{Proposition 2} 
\emph{\emph{(Chernoff-Stein's Lemma~\cite{CT2})}
Suppose that ${\rm VAR}_0[\log \left(q_i(X)/p_i(X)\right)] \le \sigma^2 < \infty$, and let 
\[ P^\ast_{n,\theta} = \displaystyle{\min_{\footnotesize{ A_n \subseteq \cX^n \, : \, \P_0(\bX^n \not \in A_n) \le \theta}}}  \P_1(\bX^n \in A_n),\]
where $0 < \theta < 1/2$.
Then
\beq
\lim_{n\rightarrow \infty} - \frac 1 n \log P^\ast_{n,\theta} = \bar{D}(q_{1:\infty}\| p_{1:\infty}).
\label{eq:stein}
\eeq}

\vspace*{2pt}\noindent\emph{Proof:} See Appendix~\ref{app:P1} for a sketch of proof.\hfill~$\bullet$

In words: 
for ``arbitrarily'' constrained
type~I error exponent, type~II error exponent can be made equal to $\bar{D}(q_{1:\infty} \| p_{1:\infty})$, but not larger.

\section{Detection with Unlabeled Data}
\label{sec:unlabeled}

Consider now the case of \emph{unlabeled} data.
Suppose that, instead of~(\ref{eq:test}), we are faced with a binary hypothesis test in which we observe the unlabeled vector $\bX^n_u \dfz \cM^{(\pi)} \bX^n$, where $\cM^{(\pi)}$ is a permutation matrix, indexed by an unknown $\pi \in \{1,\dots,n!\}$. Namely, let us consider the following test: 
\beq 
\begin{array}{c}
\bX^n_u= \cM^{(\pi)} \bX^n \; \textnormal{with} \;   \bX^n \sim r_{1:n}(\bx^n) = \prod_{i=1}^n r_i(x_i), \vspace*{5pt} \\ 
\textnormal{ where } \left \{
\begin{array}{lcl}
\cH_1 : && r_i(x_i)= p_i(x_i), \\ 
\cH_0 : && r_i(x_i)= q_i(x_i),
\end{array}
\right .
\end{array}
\label{eq:testun}
\eeq
for $n=1,2,\dots$, where the permutation matrix applied to the data is unknown.

We know that the $n$ observations are drawn from the $n$ PMFs $\{p_i\}_{i=1}^n$ under $\cH_1$
and from the $n$ PMFs $\{q_i\}_{i=1}^n$ under $\cH_0$, but we cannot make the association between observations and PMFs. In other words, 
under $\cH_1$, for each $X_j$, $j=1,\dots,n$, we do not know which, among the $n$ PMFs $\{p_i\}_{i=1}^n$ has been drawn from, and the same is true under $\cH_0$,
with $\{p_i\}_{i=1}^n$ replaced by $\{q_i\}_{i=1}^n$.

Given a constraint on type~I error, what is the best asymptotic performance in terms of exponent rate for type II error, when one has only access to the unlabeled vector $\bX^n_u$? Does there exist an equivalent of Proposition~1 for unlabeled data?
The answers are based on the following obvious but important lemma. Note that $\cI(A)$ denotes the indicator of the event $A$.

\vspace*{3pt}
\noindent
\textsc{Lemma} 
\emph{\emph{(Unlabeled Vectors and Types)}: For independent random variables drawn from a common finite alphabet~$\cX$,
knowledge of the unlabeled version $\bX^n_u$ of vector $\bX^n$ is equivalent, for the detection purposes at hand, to knowledge of the \emph{type} (or empirical PMF) of $\bX^n$, which is
\beq
t_{\bX^n}(x)\dfz\frac 1 n \sum_{i=1}^n \cI(X_i=x), \quad  x \in \cX. \label{eq:tv}
\eeq}

\vspace*{2pt}\noindent
Thus, a detection problem where the observation is the unlabeled vector $\bx^n_u$, is equivalent to a detection problem in which one observes $t_{\bx^n} \in \cP_n$, where $\cP_n$ denotes the class of $n$-types.

Detection with unlabeled data can be also regarded in the framework of invariance theory~\cite[Chap.\ 6]{Lehmann-testing3}. Under $\cH_1$ we have a class of possible  distributions because we only know that one of the $n!$ permutation matrices $\cM^{(\pi)}$ has been applied to the unobserved $\bX^n$, but we do not know which. 
For this composite hypothesis test, we can consider the class of invariant tests under the group of the $n!$ permutations of the data, which are the tests that depend on the data only through the type vector $t_{\bX^n}$, see~\cite[Th.\ 6.2.1]{Lehmann-testing3}. A UMP (uniformly most powerful) invariant test can be found as shown in~\cite[Th.\ 6.3.1]{Lehmann-testing3}, which reduces the composite problem to a simple hypothesis test. 
In the forthcoming Theorem~2 we use a different test which is easier to analyze asymptotically.

Central to our development is the function $\varphi_{\cH_h}: \Re^{|\cX|-1} \mapsto (0,\infty)$ defined, under both hypotheses $\cH_h$, $h=0,1$, as follows.
Consider the reduced alphabet $\cX^\prime \dfz \cX \setminus\{x^\prime\}$ in which an arbitrarily selected entry, say $x^\prime \in \cX$, is excluded.
Recall~from (\ref{eq:testun}) that $r_i$ denotes the distribution of the $i$-th observation. We let
\beq
\varphi_{\cH_h}(\lambda;r_i) \dfz \log   \sum_{x\in \cX} r_i(x) e^{\lambda(x)}, \label{eq:baq}
\eeq
where vector $\lambda \in \Re^{|\cX|-1}$ has entries $\lambda(x)$, $x \in \cX^\prime$, plus the \emph{dummy} entry $\lambda(x^\prime)=0$.
Clearly, $\varphi_{\cH_h}(\lambda;r_i)<\infty$ for all $\lambda\in \Re^{|\cX|-1}$ and $\varphi_{\cH_h}(0;r_i)=0$. 
In Appendix~\ref{app:proper} it is shown that $\varphi_{\cH_h}(\lambda;r_i)$ is
strictly convex and twice continuously differentiable throughout $\lambda \in \Re^{|\cX|-1}$. It is also shown that the gradient $\nabla \varphi_{\cH_h}$ is a mapping from $\lambda \in \Re^{|\cX|-1}$ to the set of $|\cX|-1$ positive values $0<\omega(x)<1$, $x\in \cX^\prime$, which, with the addition of the entry $\omega(x^\prime)=1-\sum_{x\in\cX^\prime} \omega(x)$, becomes the set of probability distributions $\omega \in \cP(\cX)$ having strictly positive entries. Henceforth, we assume that vector $\lambda(x)$, $x\in \cX^\prime$, is enlarged by the addition of $\lambda(x^\prime)=0$ and, likewise, vector $\omega(x)$, $x \in\cX^\prime$,  is enlarged by the addition of $\omega(x^\prime)$. This way, a point of the domain or range of the gradient mapping is specified by $|\cX|-1$ coordinates. 
Using this formalism, Appendix~\ref{app:proper} also shows that the gradient of (\ref{eq:baq}) evaluated at
the origin is $\nabla \varphi_{\cH_h}(0;r_i)=r_i$ (namely, equal to $p_i$ under $\cH_1$ and to $q_i$ under $\cH_0)$.

Let us introduce the arithmetic average of $\varphi_{\cH_h}(\lambda; r_i)$ over the index $i=1,2,\dots$:
\beq
\psi_{\cH_h}(\lambda;r_{1:\infty}) \dfz \lim_{n\to\infty} \frac 1 n \sum_{i=1}^n \varphi_{\cH_h}(\lambda;r_i) . \label{eq:newpsi}
\eeq

The assumption of the theorems to be presented shortly is that the aforementioned properties of $\varphi_{\cH_h}(\lambda;r_i)$, shown in Appendix~\ref{app:proper},
carry over to $\psi_{\cH_h}(\lambda;r_{1:\infty})$ after taking the arithmetic average. This is formalized in Assumption A that follows, which certainly verified in situations of interest.
One important example in which Assumption A is easily verified is when the infinite sequence of probability distributions $r_{1:\infty}$ contains only a finite number of different elements, in which case the arithmetic average
in~(\ref{eq:newpsi}) reduces to a finite sum. Note also that strict convexity of $\psi_{\cH_h}(\lambda;r_{1:\infty})$ 
always follows by the analogous property of $\varphi_{\cH_h}(\lambda;r_{i})$
because infinite positively-weighted sums of strictly convex functions preserve strict convexity~\cite{boyd-vandenberghe}.  Let $\bar r = \lim_{n\to\infty} \frac 1 n \sum_{i=1}^n r_i$ be the arithmetic average of the distributions in force.

\vspace*{3pt}
\noindent
\textsc{Assumption A.}
\emph{For $h=0,1$, function $\psi_{\cH_h}(\lambda;r_{1:\infty})$ is finite,
strictly convex and twice continuously differentiable throughout $\Re^{|\cX|-1}$, with $\psi_{\cH_h}(0; r_{1:\infty})=0$. Its gradient defines a mapping
$\nabla \psi_{\cH_h}: \Re^{|\cX|-1} \mapsto \cP(\cX)$, with $\nabla\psi_{\cH_h}(0;r_{1:\infty})=\bar r$.}

\vspace*{3pt}
\noindent
The Legendre transform of $\psi_{\cH_h}(\lambda;r_{1:\infty})$ is~\cite{rockafellar-book}:
\beq
\Psi_{\cH_h}(\omega;r_{1:\infty}) = \hspace*{-5pt} \sup_{\lambda \in \Re^{|\cX|-1}}\hspace*{-2pt} \left \{ \sum_{x\in \cX^\prime} \lambda(x) \omega(x) - \psi_{\cH_h}(\lambda; r_{1:\infty})\right \}, \label{eq:legt}
\eeq
where $\omega \in \cP(\cX)$.
In the next definition 
we use the notation $\Omega(\alpha)$ as an abbreviation  for $\Omega(\alpha;p_{1:\infty},q_{1:\infty})$.

\vspace*{5pt} \noindent 
\emph{\textsc{Definition} 
\emph{(Error Exponent for Unlabeled Data)}:
For $0<\alpha<\infty$, let:
\beq
\Omega(\alpha) \dfz  \hspace*{-5pt}\inf_{\omega \in \cP(\cX):\,\Psi_{\cH_0}(\omega; q_{1:\infty})< \alpha}  \hspace*{-2pt}\Psi_{\cH_1}(\omega ; p_{1:\infty}).
\label{eq:csi}
\eeq}

\vspace*{2pt}
\noindent
\textsc{Theorem~1}
\emph{(Properties of $\Omega(\alpha)$)}
\emph{
The error exponent $\Omega(\alpha)$ for unlabeled detection is continuous and convex for $\alpha>0$, takes
the value $\Omega(0)=\Psi_{\cH_1}(\bar q;p_{1:\infty})$ at the origin, is strictly decreasing over the interval 
$0 < \alpha < \Psi_{\cH_0}(\bar p;q_{1:\infty})$, and is identically zero 
for $\alpha\ge\Psi_{\cH_0}(\bar p; q_{1:\infty})$. In addition, for all $\alpha>0$,
\beq
\Omega(\alpha; p_{1:\infty}, q_{1:\infty}) \quad \left \{ 
\begin{array}{l} 
\le \Omega_{\rm lab}(\alpha; p_{1:\infty}, q_{1:\infty}) ,\\
\ge \Omega(\alpha; \bar p, q_{1:\infty}) ,\\
\ge \Omega(\alpha; p_{1:\infty}, \bar q) ,\\
\ge \Omega(\alpha; \bar p, \bar q) .
\end{array}
\right.
\label{eq:mi}
\eeq
When $r_{1:\infty}$ is the constant sequence $(\bar r, \bar r, \dots)$, we have $\Psi_{\cH_h}(\omega;\bar r)=D(\omega || \bar r)$, $h=0,1$, and the quantities 
in (\ref{eq:mi}) simplify accordingly. For instance:
$\Omega(\alpha; \bar p, \bar q)=\inf_{\omega \in \cP(\cX):\,D(\omega || \bar q)< \alpha} D(\omega || \bar p)$.}

\begin{figure}
\centering 
\hspace*{-20pt}\includegraphics[width =280pt]{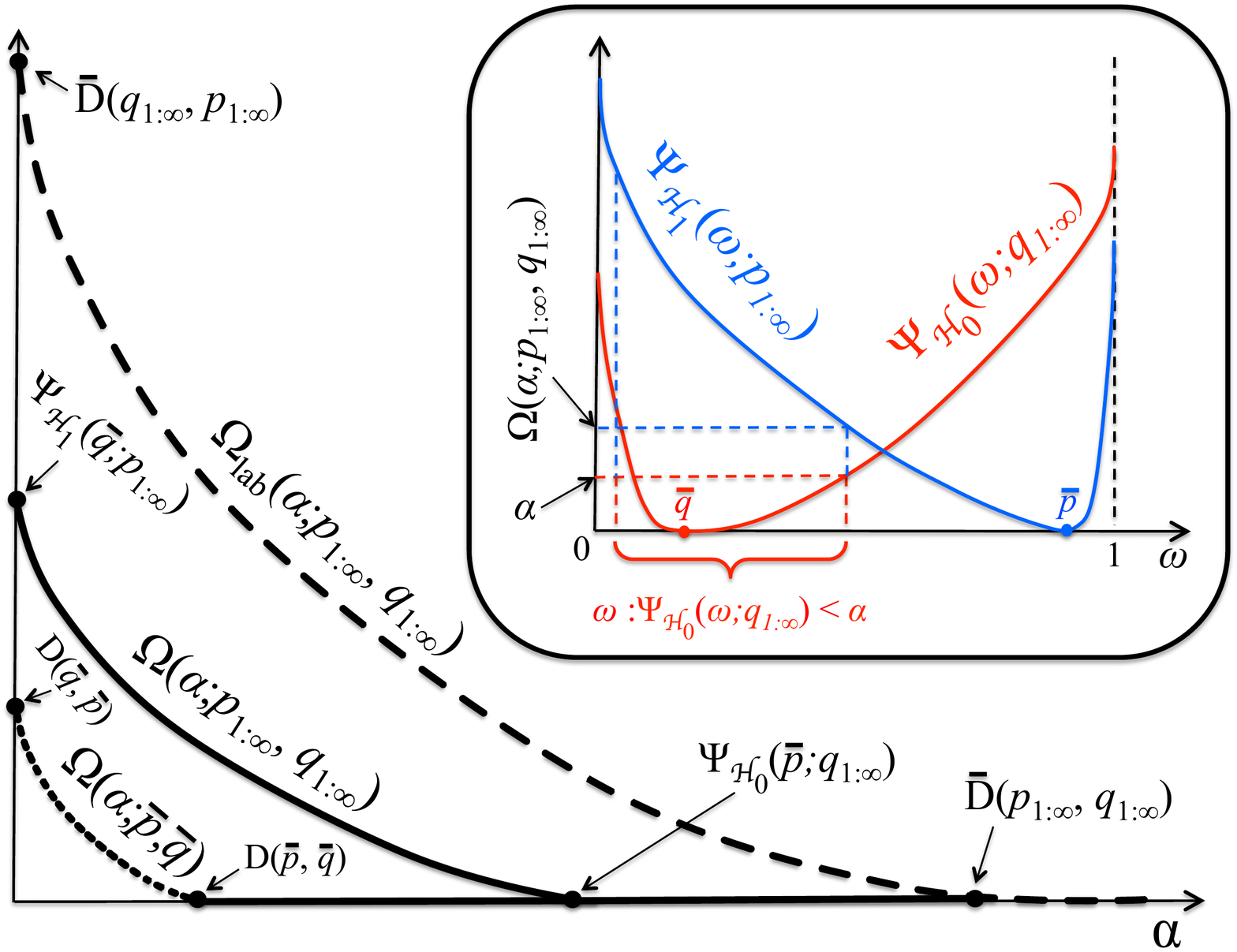}
 \caption{Error exponent for unlabeled detection $\Omega(\alpha;p_{1:\infty},q_{1:\infty})$, see~(\ref{eq:csi}). Also shown are the upper bound $\Omega_{\rm lab}(\alpha;p_{1:\infty},q_{1:\infty})$ of~(\ref{eq:Psi}),
and the lower bound $\Omega(\alpha;\bar p,\bar q)$.
The inset shows how $\Omega(\alpha;p_{1:\infty},q_{1:\infty})$ is computed 
 from $\Psi_{\cH_1}(\omega; p_{1:\infty})$ and $\Psi_{\cH_0}(\omega;q_{1:\infty})$, for the case $|\cX|=2$ with $\omega$ scalar.}
      \label{fig:intro}
 \end{figure}

\vspace*{3pt}\noindent\emph{Proof:}
The proof is given in Appendix~\ref{app:proper}.\hfill~$\bullet$\vspace*{3pt}

Our main theoretical result is contained in the following theorem, which provides the operational meaning of $\Omega(\alpha)$ and extends Proposition~1 to unlabeled detection.

\vspace*{3pt}
\noindent
\textsc{Theorem~2}
\emph{\emph{(Unlabeled Detection)} Consider the hypothesis test with unlabeled data formalized in~(\ref{eq:testun}). 
Suppose that Assumption A is verified, and let $0<\alpha<\infty$. \\
\noindent 
a) For any closed acceptance region $E \subseteq \cP(\cX)$ for~$\cH_0$:
\begin{align}
&\liminf_{n\to\infty} - \frac 1 n \log \P_0(t_{\bX^n} \not \in E) \ge \alpha \nonumber \\
& \;  \Rightarrow \;  \limsup_{n\to\infty} - \frac 1 n \log \P_1(t_{\bX^n} \in E) \le \Omega(\alpha)  . \label{eq:ub}
\end{align} 
b) Setting $E^\ast=\{\omega \in \cP(\cX): \Psi_{\cH_0}(\omega;q_{1:\infty})\le\alpha\}$, we get 
\begin{subequations}\begin{align}
&\liminf_{n\to\infty} - \frac 1 n \log \P_0(t_{\bX^n} \not \in E^\ast) \ge \alpha \label{eq:t3a} \\
&  \lim_{n\to\infty} - \frac 1 n \log \P_1(t_{\bX^n} \in  E^\ast) = \Omega(\alpha) . \label{eq:t3b}
\end{align}\end{subequations}}

\vspace*{2pt}\noindent\emph{Proof:}
The proof is given in Appendix~\ref{app:P2}.\hfill~$\bullet$\vspace*{3pt}

Note that the asymptotically optimal region of part b) does not require  knowledge of the sequence $p_{1:\infty}$. 
The interpretation of Theorem~2 is similar to the interpretation of Proposition~1: 
With unlabeled data it is possible to find tests such that type~I error is $\stackrel{\cdot}{=} e^{-n \alpha}$ and  type~II error is $ \stackrel{\cdot}{=}$ $e^{-n \Omega(\alpha)}$,
but no stronger pairs of asymptotic expressions can be simultaneously achieved. 
Figure~\ref{fig:intro} depicts the typical behavior of the error exponent $\Omega(\alpha)$. 

Note by Theorem~1 that $\Omega(\alpha)$ is upper bounded by the error exponent for labeled data, and lower bounded by the exponent obtained when data under either (or both) hypotheses  are drawn \emph{iid} according to the average distributions $\bar p$ or $\bar q$.
The upper and lower bounds in $\Omega(\alpha;\bar p, \bar q) \le \Omega(\alpha; p_{1:\infty}, q_{1:\infty}) \le \Omega_{\rm lab}(\alpha; p_{1:\infty}, q_{1:\infty})$ coincide when data are \emph{iid} under both hypotheses, as it must be.

\begin{figure}
\centering 
\includegraphics[width =220pt]{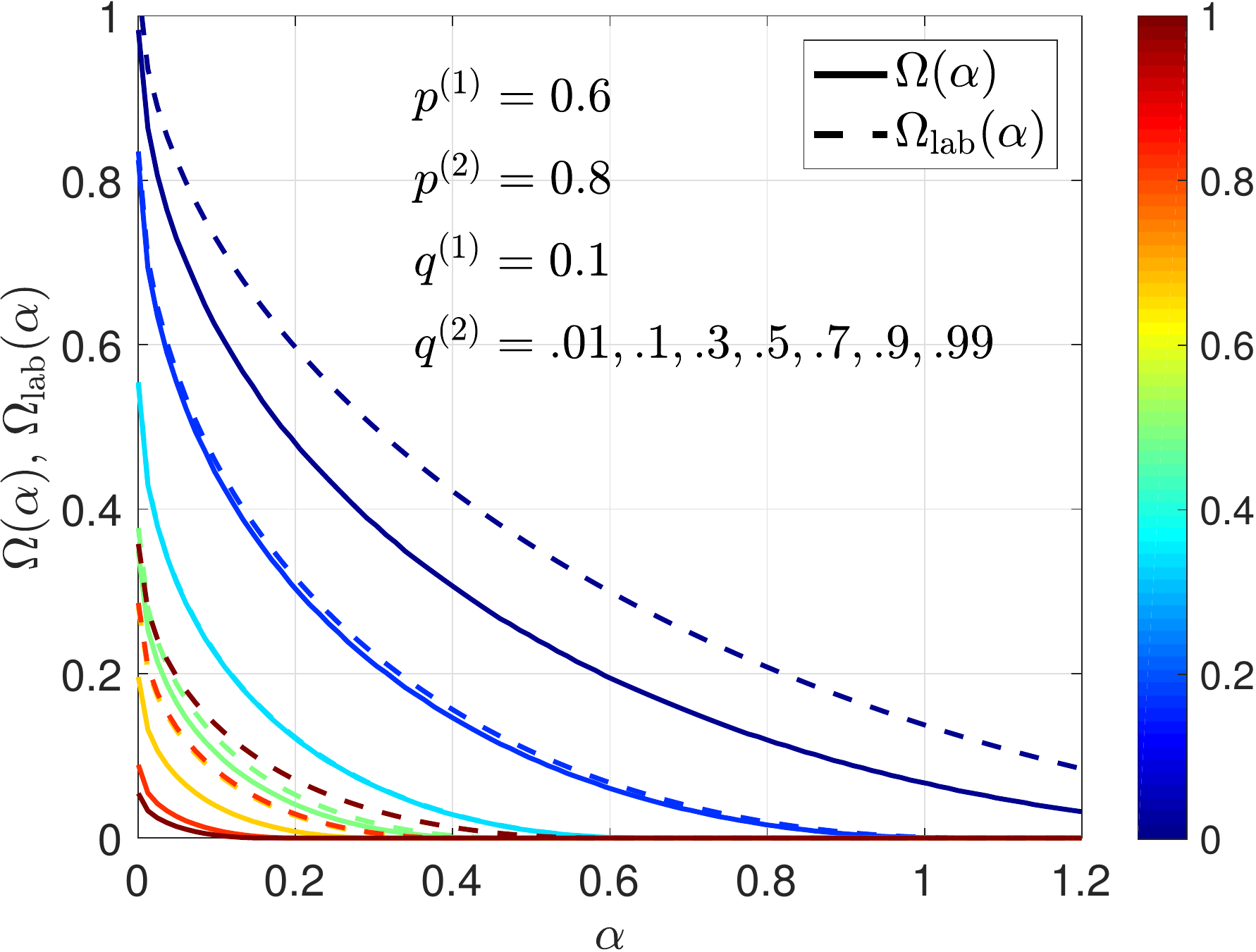}
 \caption{Error exponents for labeled and unlabeled detection in the case of half-and-half binary observations described in the main text.}
      \label{fig:1bis}
 \end{figure}

As an example of application of the theorem, let us consider the binary case $|\cX|=2$, and suppose that under $\cH_1$ half observations
are drawn from distribution $(p^{(1)},1-p^{(1)})^T$ and half from $(p^{(2)},1-p^{(2)})^T$, where $^T$ denotes vector transposition. Likewise, under $\cH_0$ half observations are drawn from distribution $(q^{(1)},1-q^{(1)})^T$ and half from $(q^{(2)},1-q^{(2)})^T$. 
In this case the divergence rates appearing in definition~(\ref{eq:Psi}) reduce to the balanced sum of only two divergences, and the infimum in~(\ref{eq:Psi}) is computed over the set $\cP(\cX^2)$.
The error exponents $\Omega(\alpha)$ and $\Omega_{\rm lab}(\alpha)$ for this detection problem are depicted in Fig.~\ref{fig:1bis}, where different values of $q^{(2)}$ are shown with the colors indicated by the color bar. Note that there exist combinations of the parameters for which $\Omega(\alpha)$ and $\Omega_{\rm lab}(\alpha)$ are very close to each other, and there exist combinations for which the information contained in the labels is very relevant and $\Omega(\alpha)$ is substantially smaller than $\Omega_{\rm lab}(\alpha)$.
The extreme case $\Omega(\alpha)=\Omega_{\rm lab}(\alpha)$ is also possible. Aside from the obvious \emph{iid} case $p^{(1)}=p^{(2)}$ and $q^{(1)}=q^{(2)}$, this happens when $p^{(1)}+q^{(2)}=p^{(2)}+q^{(1)}=1$, which can be explained by noting that the corresponding log-likelihood ratio is a function of the \emph{type} of the observed vector, 
and therefore the optimal unlabeled detector performs as the optimal labeled one.

\section{Practical Algorithms for Unlabeled Detection}
\label{sec:twodet}

Part b) of Theorem~2 gives an explicit expression of the acceptance region of the optimal test. However, this leaves open many practical questions. First, the optimality of the test shown in Theorem~2 is only asymptotic and little can be said on its performance for finite --- possibly ``small'' --- values of~$n$. Second, more important, no attention has been paid to the computational complexity required to implement the test. Third, in some applications it is desirable to recover the lost labels. These practical aspects are now addressed by considering specific detectors.

A first detector is introduced by making an analogy with 
the following detection problem with \emph{labeled} data: $\cH_1 \hspace*{-4pt}: \widetilde \bX^n \sim \bar p_{1:\infty}=(\bar p,\bar p,\dots)$ versus $\cH_0 \hspace*{-4pt}: \widetilde \bX^n \sim \bar q_{1:\infty}=(\bar q,\bar q,\dots)$, where the entries of $\widetilde \bX^n$ are now \emph{iid} under both hypotheses. The optimal decision statistic for this test is the log-likelihood ratio $\sum_{x \in\cX} n t_{\widetilde \bx^n}(x) \log \frac{\bar p(x)}{\bar q(x)}$. 
For large~$n$, $t_{\widetilde \bx^n} \approx t_{\bx^n}$, in the sense shown in Appendix~\ref{app:wp1}. We then propose the following detection statistic for unlabeled data:
\beq
\sum_{x \in\cX} t_{\bx^n}(x) \log  \frac{\bar p(x)}{\bar q(x)},
\eeq
which is referred to as the statistic of the unlabeled log-likelihood ratio (ULR) detector. 
Were $t_{\widetilde \bx^n}$ equal to $t_{\bx^n}$ the error exponent of the test would be $\Omega(\alpha; \bar p, \bar q)$, which is only a lower bound to the optimal performance of unlabeled detection, as shown by Theorem 2. However closeness of $t_{\bx^n}$ to $t_{\widetilde \bx^n}$ tells nothing about the rate of convergence to zero of the detection errors, and nothing can be anticipated as to the performance of this detector. Its main advantage is its low computational complexity:
With the type vector $t_{\bx^n}$ available, its implementation only requires $|\cX|$ multiplications and $|\cX|-1$ additions, independently of $n$: the complexity is~${\cal O}(1)$.

The ULR detector makes no attempt to estimate the labels. When an estimate of the lost labels is required, a different approach must be pursued. To elaborate,
let $\cX=\{1,2,\dots,m\}$ be the observation alphabet, which entails no loss of generality, and let us start from the case in which the detector observes the \emph{labeled} vector $\bx^n=(x_1,\dots,x_n)$, see~(\ref{eq:test}). Let $\log p_i(k)- \log q_i(k) \dfz u_{ki} -v_{ki}$ be the marginal log-likelihood ratio of the $i$-th observed sample $x_i$, when $x_i=k$, $k=1,\dots,m$. Organizing these values in $m$-by-$n$ matrix form, we have:
\beq
\small
\left (
\begin{matrix}
u_{11}-v_{11}  &         \pmb{u_{12}-v_{12}}  &         u_{13}-v_{13}   &        \dots      &  u_{1n}-v_{1n}  \\  
u_{21}-v_{21}  &         u_{22}-v_{22}  &         u_{23}-v_{23}   &        \dots      &    u_{2n}-v_{2n}  \\  
\pmb{u_{31}-v_{31}}  &         u_{32}-v_{32}  &         u_{33}-v_{33}   &        \dots       & \pmb{u_{3n}-v_{3n}}  \\  
    \vdots &       \vdots   &    \vdots      &         \ddots       &     \vdots         \\
u_{m1}-v_{m1}  &         u_{m2}-v_{m2}  &         \pmb{u_{m3}-v_{m3}}   &        \dots        & u_{mn}-v_{mn}
\end{matrix}
\right ).
\label{eq:matrix}
\eeq
The optimal log-likelihood ratio statistic for test~(\ref{eq:test}) is ${\cal U}-{\cal V}$, where ${\cal U}=\sum_{i=1}^n u_{k_ii}$ and ${\cal V}=\sum_{i=1}^n v_{k_ii}$, with $k_i$ denoting the value taken by the $i$-th observation $x_i$. 
The statistic ${\cal U}-{\cal V}$ involves $n$ entries of matrix~(\ref{eq:matrix}). Precisely, one entry over each column and $n \, t_{\bx^n}(k)$ entries over the $k$-th row. 
In other words, regarding the above matrix as a trellis (left to right), the optimal log-likelihood statistic for test~(\ref{eq:test}) is obtained by summing the entries belonging to a specific \emph{path} over the trellis~(\ref{eq:matrix}). 
For instance, if the observed vector is $\bx^n=(3,  1,  m,  \dots,  3)$, the optimal path is that shown in~(\ref{eq:matrix}) by boldface symbols.

The point with unlabeled detection is that we do not observe $\bx^n$ but only its type $t_{\bx^n}$,
and the optimal path across the trellis in unknown.  
Note that the ``optimal'' test (Bayesian, assuming that all permutations are equally likely) is the ratio of two averaged likelihoods.
One is sum of the likelihoods for $\cH_1$ over all possible permutations of labels, divided by the number of permutations, and the other is
the analogous average of the likelihoods for $\cH_0$. That this Bayesian test is infeasible, even with the simplification of considering the types, for any reasonable size of problem is self evident. 

One possible approach to circumvent the lack of precise knowledge of the optimal path is to resort to the 
generalized likelihood ratio test (GLRT). The GLRT consists of replacing the unknown labeling by its maximum likelihood estimate under each hypothesis,
and then constructing the ratio of the resulting likelihoods~\cite{kaydetection}.
The GLRT is not an optimal test but in many instances may led to nicely-performing tests amenable of simple implementation, and gives us as by-product an estimate of the permutation under $\cH_1$ and under $\cH_0$. Thus, after the decision about the hypothesis is made, the pertinent estimate of the labeling is also available.

Returning to the GLRT, to see how it works consider first the log-likelihood for $\cH_1$, represented by the analogous of matrix~(\ref{eq:matrix}) containing only the values $\{u_{ki}\}$. Among all the possible paths across such trellis, the GLRT selects the one yielding the largest sum among all paths compatible with the observed $t_{\bx^n}$. The compatible paths are those with one entry per column, and $n \, t_{\bx^n}(k)$ entries over the $k$-row.
A convenient way to visualize these paths is to introduce an \emph{augmented} version of the trellis, where the $k$-th row of~(\ref{eq:matrix}) is copied 
$n \, t_{\bx^n}(k)$ times, for $k=1,\dots,n$. This yields the following $n$ by $n$ trellis:
\beq
\footnotesize
\begin{array}{cccccccc}
\multirow{3}{*}{$n \, t_{\bx^n}(1)$  copies \Bigg \{ }  & \hspace*{-22pt}  \multirow{11}{*}{$\left (\begin{array}{c} \\ \\ \\ \\ \\ \\ \\ \\ \\ \\ \\ \end{array} \right .$ \hspace*{-50pt}  }  &
u_{11}  &         u_{12}  &    u_{13}  &         \cdots      &    u_{1n} & \hspace*{-18pt}\multirow{10}{*}{ $\left . \begin{array}{c} \\ \\ \\ \\ \\ \\ \\ \\ \\ \\ \\ \end{array} \right )$ } \\ 
& & \cdots &  \cdots &  \cdots & \cdots & \cdots \\  
& & u_{11}  &         u_{12}  &  u_{13}  &              \dots      &     u_{1n} \\ 
\multirow{3}{*}{$n \, t_{\bx^n}(2)$  copies \Bigg \{ } & &
u_{21}  &         u_{22}  &  u_{23}  &           \cdots      &     u_{2n}  \\ 
& & \cdots &  \cdots &  \cdots & \cdots & \cdots \\  
&& u_{21}  &         u_{22}  &    u_{23}  &            \dots      &     u_{2n} \\ 
\vdots  &&\vdots & & \vdots & \ddots & \vdots\\ 
\multirow{3}{*}{$n \, t_{\bx^n}(m)$ copies \Bigg \{ }  & &
u_{m1}  &         u_{m2}  &      u_{m3}  &        \cdots      &     u_{mn}  \\ 
& & \cdots &  \cdots &  \cdots & \cdots & \cdots \\  
& & u_{m1}  &         u_{m2}  &       u_{m3}  &      \dots      &     u_{mn} 
\end{array} 
\label{eq:aut}
\eeq

Finding the ``GLRT path'' across the augmented trellis~(\ref{eq:aut}) amounts to select one entry over each row and one entry over each column, 
with the goal of maximizing the sum of the $n$ selected entries. Let us denote by ${\cal U}_{\rm GLRT}$ this maximum sum.
Likewise, for $\cH_0$, we consider the trellis similar to~(\ref{eq:aut}) with the $u_{ki}$'s replaced by the $v_{ki}$'s. The best path over this new trellis
must be found\footnote{Of course, the orderings may (and likely will) be completely different different under the two hypotheses.}, and the sum of the corresponding entries is denoted by ${\cal V}_{\rm GLRT}$. The GLRT statistic is ${\cal U}_{\rm GLRT}-{\cal V}_{\rm GLRT}$, and requires to find two optimal paths, which represent the estimate of the labels under the two hypotheses.

Finding the best path over these trellises is not a combinatorial problem, because exhaustive search is not necessary.
Indeed, the search of the GLRT path across a trellis like that in~(\ref{eq:aut}) is an instance of the transportation problem --- a special case of the assignment problem --- for which efficient algorithms have been developed~\cite{bertsekas-tsitsiklis}. In the jargon of the assignment problem, each row of~(\ref{eq:aut}) represents a ``person'', each column represents an ``object'', and the $(k,i)$-th entry is the benefit for person~$k$ if obtains object~$i$. The problem is to assign one distinct object to each person providing the maximum global benefit.

The Hungarian (a.k.a.\ Munkres or Munkres-Kuhn) algorithm solves exactly the assignment problem in ${\cal O}(n^3)$ operations~\cite{Kuhn-Hungarian,Munkres}. 
The auction method usually has lower complexity and is amenable to parallel implementation. A nice overview of the auction procedure and its application to data association can be found in \cite{blackman-popoli}. Among the many variants of the auction method, the $\epsilon$-scaled implementation achieves a solution of the assignment problem $n \epsilon$-close to the actual maximum~\cite{bertsekas-netopt}. The computational complexity of the auction algorithm depends on the data structure and when the assignment problem involves \emph{similar persons} (i.e., equal rows, as in our case) it can be highly inefficient~\cite{bertsekas-castanon}. 
A variation of the auction algorithm specifically tailored to address assignment problems with similar persons and similar objects has been proposed in~\cite{bertsekas-castanon,JVC}. The auction algorithm used in this paper is $\epsilon$-scaled and is a special form of that proposed in~\cite{bertsekas-castanon}, accounting for the presence of similar persons but \emph{not} of similar objects. This algorithm is here referred to as ``auction-sp''.

Aside from the auction-sp algorithm, we present two greedy procedures. These detection algorithms are easily described by referring to a simple example.
Suppose we have $n=5$ observations, and assume that the alphabet is $\cX=\{1,2,3\}$.  
Consider the following trellis whose $(k,i)$-th entry is $u_{ki}=\log p_i(k)$, $i=1,\dots,n$, $k=1,\dots, m$:
\beq
\left (
\begin{matrix}
       \log(1/10)     &      \log(1/12)     &       \log(1/6)    &        \pmb{\log(1/4)}     &       \pmb{\log(1/3)}    \\  
       \log(3/10)      &     \pmb{\log(1/3)}        &      \log(1/3)       &       \log(1/3)       &       \log(1/3)      \\ 
       \pmb{\log(3/5)}       &      \log(7/12)       &      \pmb{\log(1/2)}     &        \log(5/12)     &       \log(1/3)
       \end{matrix}
\right ).
\label{eq:LLM}
\eeq
Finally, suppose that the vector of (labeled) observations is $\bx^5=(3 \;2 \;1 \;3 \;1)$. 
With unlabeled data, vector $\bx^5 $ is not available and we only observe the type $t_{\bx^5}=(2\;1\;2)$ or, what is the same, the \emph{sorted} version of $\bx^5$, namely $\bx_{\rm sort}^5=(1\;1\;2\;3\;3)$.

\subsection{Detector A}

With reference to the above example, the first algorithm we propose processes sequentially the entries of $\bx_{\rm sort}^5=(1\;1\;2\;3\;3)$ and selects, for each entry, the step (column) on the trellis~(\ref{eq:LLM}) with largest value. 
For instance, consider the first entry of~$\bx_{\rm sort}^5$, which is~1. By inspection of the trellis~(\ref{eq:LLM}) 
we see that the maximum value attained by the first row of the matrix is $\log(1/3)$
and is attained at the fifth column. Therefore, we assign the state~1 to the fifth step of the trellis. At this point, the fifth column of matrix~(\ref{eq:LLM})  is blocked and excluded from the analysis, and we move to consider the second element of $\bx_{\rm sort}^5$, whose value is again 1. We inspect again the first row of matrix~(\ref{eq:LLM}), ignoring its fifth entry. 
The maximum value $\log(1/4)$  is attained at the fourth column and therefore we assign the state 1 to the fourth step of the path. Next, consider the third entry of $\bx_{\rm sort}^5$, whose value is~2, and consider the second row of matrix~(\ref{eq:LLM}), ignoring its fourth and fifth entries. The maximum is $\log(1/3)$ and attained at the second and at the third column. In the case of ties, an arbitrary choice is made: We choose the former, and the state~2 is assigned to the $2$-nd step of the path.  We consider now the fourth entry of $\bx_{\rm sort}^5$, which is~3, and inspect the third row of matrix~(\ref{eq:LLM}), ignoring its second, fourth, and fifth entries. The largest between the first and the third entries of the third row of~(\ref{eq:LLM}) is attained at the former, which implies that the state 3 is assigned to the first step of the path. We have been left with the last entry of $\bx_{\rm sort}^5$, which is~3, and the only surviving column of matrix~(\ref{eq:LLM}) is the third: the state~3 is assigned to the third step of the path. We have arrived at determining the path $(3\;2\;3\;1\;1)$, which is emphasized in bold in~(\ref{eq:LLM}).
The first contribution ${\cal U}_{\rm A}$ to the decision statistic for Detector~A is the sum of the entries in bold. By repeating the path search over the trellis 
similar to~(\ref{eq:LLM}) but with the $u_{ki}$'s replaced by the $v_{ki}$'s, we obtain the second contribution ${\cal V}_{\rm A}$, and the 
decision statistic for Detector~A is given by the difference ${\cal U}_{\rm A}-{\cal V}_{\rm A}$.

The computational cost of Algorithm~A can be approximately evaluated by noting that the $k$-th iteration 
amounts to computing the maximum of a $(n-k+1)$-sized vector of reals, and there are $n-1 \approx n$ such iterations. 
If we assume that computing the maximum over $\ell$ numbers requires a number of elementary operations proportional to $\ell$, an approximate value for the computational cost is proportional to
$\sum_{k=1}^n (n-k+1)$ $=$ $n(n+1)/2$,
namely, the computational complexity of Algorithm~A is~${\cal O}(n^2)$.

\vspace*{-2pt}
\subsection{Detector B}

Consider again the trellis~(\ref{eq:LLM}) and the unlabeled vector $\bx_{\rm sort}^5=(1\;1\;2\;3\;3)$. Algorithm~B works as follows. First, regardless of the observed 
vector $\bx_{\rm sort}^5$, we select the best path on the trellis, in the sense of achieving the largest sum of $n$ entries, choosing one entry per column.
This yields the path shown below in bold:
\beq
\left (
\begin{matrix}
       \log(1/10)     &      \log(1/12)     &       \log(1/6)    &        \log(1/4)     &       \pmb{\log(1/3)}    \\  
       \log(3/10)      &     \log(1/3)        &      \log(1/3)       &       \log(1/3)       &       \log(1/3)      \\ 
       \pmb{\log(3/5)}       &      \pmb{\log(7/12)}       &      \pmb{\log(1/2)}     &        \pmb{\log(5/12)}     &       \log(1/3)
       \end{matrix}
\right )
\label{eq:LLMo}
\eeq
where in the last column any entry could be chosen, and we arbitrarily select the first. Should the observed unlabeled vector have been $(1\;3\;3\;3\;3)$, the above largest-value path would be compatible with the observations, but it is not so. Algorithm~B now proceeds to make the minimum number of modifications to the path in bold in~(\ref{eq:LLMo}), up to obtain a path compatible with the observed $\bx_{\rm sort}^5=(1\;1\;2\;3\;3)$. 
By comparing the path $(3\;3\;3\;3\;1)$ in~(\ref{eq:LLMo}) to the observed $(1\;1\;2\;3\;3)$, we see that the path $(3\;3\;3\;3\;1)$ requires two modifications, and in particular two states with value 3 must be modified to become 2 and 1, respectively. In symbols $(3\;3) \mapsto (1\;2)$. Let us address these modifications sequentially.

Thus, consider the first modification $3 \mapsto 1$. The path in~(\ref{eq:LLMo}) has state 3 in correspondence of the first four steps, and we have to choose which of these steps we want to modify the state  from 3 to 1. The most appropriate choice is to make the change $3 \mapsto 1$ in correspondence of the fourth step, because this modification reduces the total statistic [the sum of the entries emphasized in bold in~(\ref{eq:LLMo})] the minimal amount, as seen by considering the four differences $\log(3/5)-\log(1/10)$, $\log(7/12)-\log(1/12)$, $\log(1/2)-\log(1/6)$, $\log(5/12)-\log(1/4)$, which take the minimum value in the last case. Implementing this change of state yields the path $(3\;3\;3\;1\;1)$, and the fourth step in~(\ref{eq:LLMo}), where we have made a path modification, is now blocked and further modifications to it are inhibited.

We are left with one more change $3 \mapsto 2$, and the candidate path steps for such modification are the steps 1, 2 and 3, whose state is 3. Consider hence the differences $\log(3/5)-\log(3/10)$, $\log(7/12)-\log(1/3)$, $\log(1/2)-\log(1/3)$, and note that the last difference, corresponding to the third step of the path, is the smallest. Accordingly, the path $(3\;3\;3\;1\;1)$ is modified to $(3\;3\;2\;1\;1)$, which is the final path on the trellis according to Algorithm~B. The first contribution to the decision statistic 
of Detector~B is ${\cal U}_{\rm B}=\log(3/5)+\log(7/12)+\log(1/3)+\log(1/4)+\log(1/3)$. Running Algorithm~B over the trellis with entries $\{v_{ki}\}$ gives the 
contribution ${\cal V}_{\rm B}$, and the final decision statistic
for Detector~B is ${\cal U}_{\rm B}-{\cal V}_{\rm B}$. The Matlab$^{\copyright}$-style code shown below gives the general form of the path search for Algorithm~B.

\begin{algorithm}[h]
\small{
\NoCaptionOfAlgo
\DontPrintSemicolon
\KwIn{$L$,  $x$ \\ 
$\quad$ $L$: $m$-by-$n$ matrix of log-likelihood values \\ $\qquad$($m=$ alphabet size, $n=$ No.\ of samples)\\
$\quad$  $x$: $1$-by-$n$ (sorted) vector with entries $\in \{1,\dots,m\}$}
\KwOut{$p$ $\,\,\,$ path over the trellis $L$} 
\vspace*{10pt}
\textbf{function} $p = \textnormal{algorithmB}(L, x)$ \\
$[m,n]=\textnormal{size}(L)$; \\
$[d,p]=\max(L,[\,],1)$; \\ 
$sx = \textnormal{sort}(x)$; \\
$sp = \textnormal{sort}(p)$; \\
$g = \textnormal{not}(sp == sx)$; \\ 
$ch = [sp(g); sx(g)]$; \\ 
$[d, j ] = \textnormal{size}(ch)$; \\
$bl = \textnormal{false}(1, n )$; \\
\textbf{for } $i=1:j$ \\
$\quad g = \textnormal{find}((p == ch(1, i)) \, \& \, \textnormal{not}(bl))$; \\
$\quad    [d, k] = \min(L(ch(1, i)*\textnormal{ones}(\textnormal{size}(g)) + m *(g-1)) - ...$ \\
$\qquad\qquad        L(ch(2, i) * \textnormal{ones}(\textnormal{size}(g)) + m * (g-1)), [ \, ], 2)$; \\ 
$\quad    p(g(k)) = ch(2, i)$; \\
$\quad    bl(g(k)) = \textnormal{true}$; \\
\textbf{end} \\
\caption{\textbf{Algorithm B}}
}
\end{algorithm}

The computational complexity of Algorithm~B can be estimated by considering that the ``for'' cycle is the part of the routine that essentially determines the computational cost.
In this cycle the minimum over a decreasing-size vector is computed. In the worst case where all the $n$ states of the initial path must be changed, such vector has size $n$, and the same argument used for Algorithm~A leads to the conclusion that the computational complexity of Algorithm~B is ${\cal O}(n^2)$. 
However, the actual number of modifications required is less (and possibly much less) than~$n$, and depends on the realization of $\bx^n_{\rm sort}$
and on the trellis values. This implies that the computational complexity of Algorithm~B is only upper bounded by ${\cal O}(n^2)$, but can be substantially less.

\begin{figure}
\centering 
\includegraphics[width =200pt]{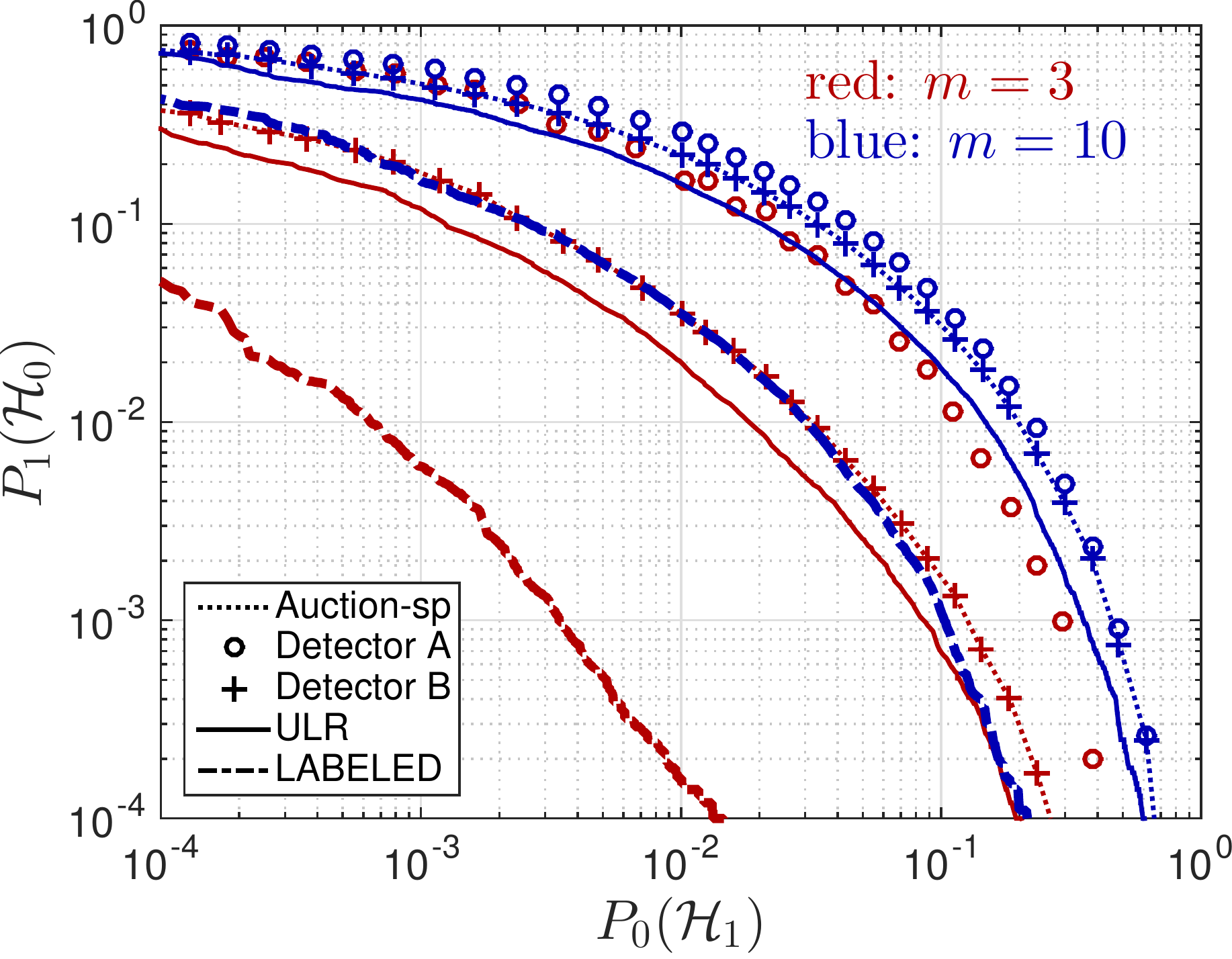}
 \caption{First computer experiment. Type~II error probability versus type~I error probability, for $n=100$ and $m=3,10$.}
      \label{fig:1}
 \end{figure}

\section{Computer Experiments}
\label{sec:compe}

Let us begin by assuming that data are \emph{iid} under $\cH_0$, so that the path search must be performed on a single trellis.
In the first computer experiment we assume  that under $\cH_0$ data are uniformly distributed, namely, $q_i=(1/m \dots 1/m)^T$, for all $i=1,\dots,n$,
and under $\cH_1$ the $n$ PMFs of size~$m$, written as columns of an $m$-by-$n$ matrix, are as follows: 
\beq
\left ( \hspace*{-3pt}
\begin{smallmatrix}
       0     & &    \frac{1/m}{n-1}     & &    2 \frac{1/m}{n-1}     &           &  &      \frac 1 m      \\  
       \kappa      &  &  \kappa+\frac{1/m-\kappa}{n-1}        & &     \kappa+2\frac{1/m-\kappa}{n-1}       &       \cdots        &&       \frac 1 m      \\ 
       2 \kappa       & &      2\kappa+\frac{1/m-2\kappa}{n-1}     &  &       2\kappa+2\frac{1/m-2\kappa}{n-1}     &         \cdots  & &       \frac 1 m   \\
        3 \kappa       & &      3\kappa+\frac{1/m-3\kappa}{n-1}     &  &       3\kappa+2\frac{1/m-3\kappa}{n-1}     &         \cdots  & &       \frac 1 m   \\
       \vdots       & &      \vdots     & &       \vdots     &       \ddots       & &     \vdots   \\ \\
       (m-1) \kappa       & \,\,\, &    (m-1) \kappa+\frac{1/m-(m-1)\kappa}{n-1}    &  \,\,\, &       (m-1) \kappa+2\frac{1/m-(m-1)\kappa}{n-1}     &     \cdots       & &       \frac 1 m   
       \end{smallmatrix}
\hspace*{-3pt} \right ),
\label{eq:matex}
\eeq
where $\kappa=\frac{2}{m(m-1)}$. Thus, the first PMF (leftmost column) is\footnote{Note that we have always assumed strictly positive PMFs. Then, for the sake of rigor, we could replace the zero in~(\ref{eq:matex}) with a sufficiently small positive value, and then normalizing to unit the first column. This removes the zero and leaves essentially unchanged the arguments and the results that follow.} 
$p_1=\begin{smallmatrix} \left (0 \,\frac{2}{m(m-1)} \, \frac{4}{m(m-1)} \dots \frac{2}{m} \right )^T \end{smallmatrix}$, the $n$-th PMF (rightmost) $p_n=\begin{smallmatrix} \left (\frac 1 m \frac 1 m \dots \frac 1 m \right )^T \end{smallmatrix}$ is uniform, and all other columns of~(\ref{eq:matex}) are such that the entries on each row vary linearly from the leftmost to the rightmost value (i.e., increase or decrease linearly). 
Straightforward calculation shows that the entries of the $n$-averaged PMF $\frac 1 n \sum_{i=1}^n p_i$ are 
\beq
\begin{matrix} \left (
\frac{1}{2m} \,\, \frac{m+1}{2m(m-1)} \,\, \frac{m+3}{2m(m-1)} \,\, \frac{m+5}{2m(m-1)} \dots \frac{m+2(m-1)-1)}{2m(m-1)}
\right)^T \hspace*{-6pt}.
\end{matrix}
\label{eq:bapex}
\eeq
Since these values do not depend on~$n$, we have that $\bar p =\lim_{n\to\infty} \frac 1 n \sum_{i=1}^n p_i$ is given by~(\ref{eq:bapex}).

 \begin{figure}
\centering 
\includegraphics[width =200pt]{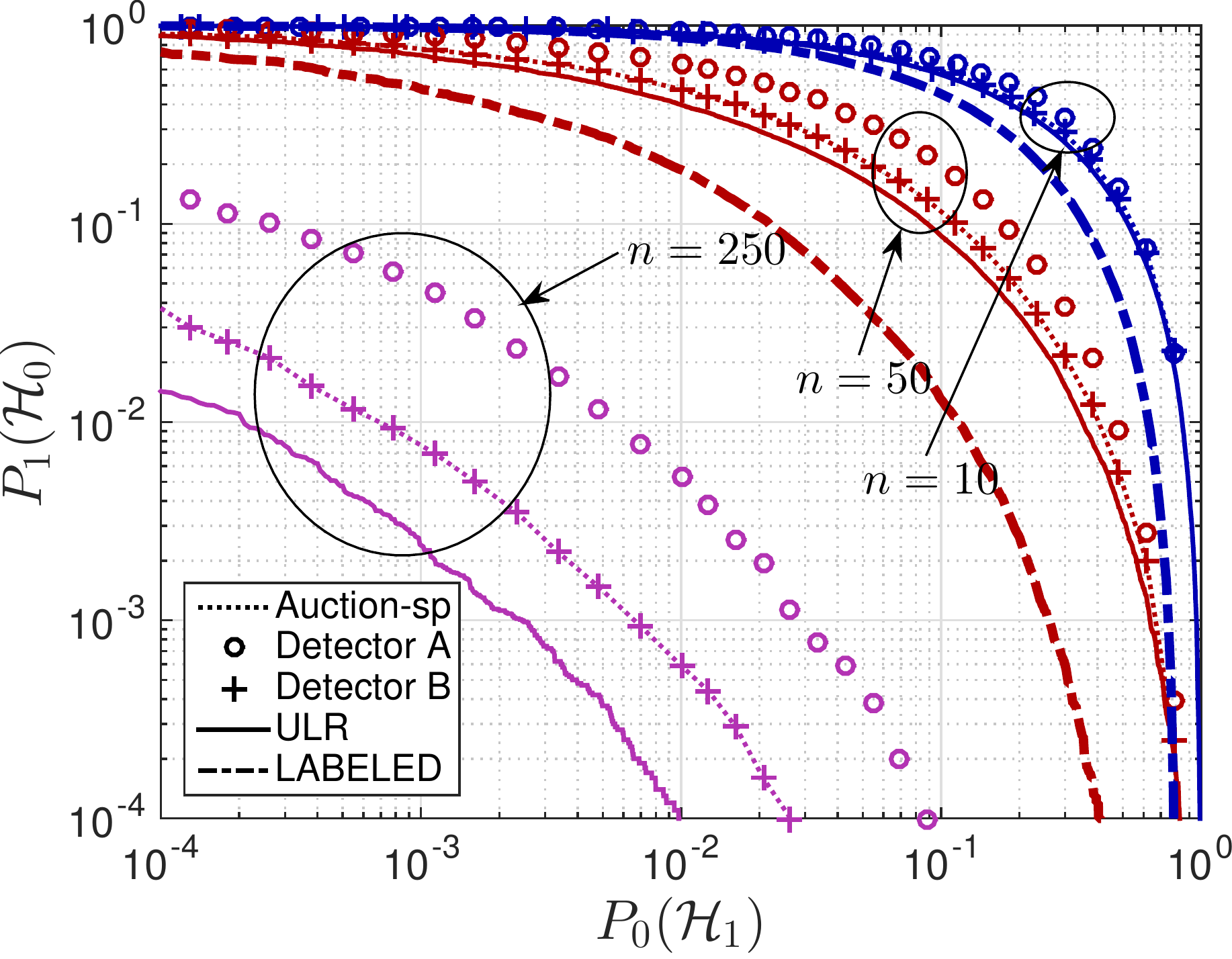}
 \caption{First computer experiment. Type~II error probability versus type~I error probability for $m=5$ and $n=10,50,250$.}
      \label{fig:2}
 \end{figure}

For this case study, we now investigate the performance of the four detectors presented in Sect.~\ref{sec:twodet}: ULR, auction-sp, detector~A, and detector~B.
For the auction-sp algorithm, after trials and errors we found that $\epsilon = 10^{-3}/m$ practically achieves the same total benefit as the Hungarian algorithm, and this value of~$\epsilon$ is therefore selected in all numerical experiments.

In Figs.~\ref{fig:1} and~\ref{fig:2}  we show the ROC (Receiver Operational Characteristic), namely the type II error versus the type I error,\footnote{Actually, the ``ROC'' curve is the complement of type II error in function of type I error.} obtained by $10^5$ Monte Carlo simulations. Clearly, the lower is the ROC curve, the better is the detection performance.
In Fig.~\ref{fig:1} we set $n=100$, and consider two values of the alphabet size $m=3,10$. For $m=3$ we see that detector~B outperforms detector~A,
detector~B performs exactly as auction-sp, and their performance is close to that of the ULR, which gives the best performance.
For the sake of comparison, we also report the ROC curve for the ``labeled'' detector, namely, for the case in which the association between data and generating PMFs is perfectly known (no data permutation takes place). As it must be, the labeled detector achieves much better performance.
Next, looking at the case $m=10$ in In Fig.~\ref{fig:1}, we see that  the performance of the detectors worsen, and their relative ordering is as for $m=3$, with a minor gain of detector~B over detector~A, and also a minor gain of the labeled detector over the unlabeled ones.

A similar analysis is carried out in Fig.~\ref{fig:2}, where $m=5$ and three values of $n$ are considered. We see that by increasing~$n$ the detection performance improves.
The performance of the labeled detector for $n=250$ is not shown because its performance is much better, and $\P_0(\cH_1)$, $\P_1(\cH_0)$, fall out of the axis range.

We now consider the computational complexity of the detection algorithms for the addressed example. The precise evaluation of the program execution time is, of course, highly dependent on a number of factors related to the specific hardware and software, and therefore would be of limited interest. 
We instead report in Table~\ref{tab:1} the ratio between the execution times of the different detectors, which is expected to be less machine/software-dependent.
The data in Table~\ref{tab:1} have been obtained by averaging the results of 1000 Monte Carlo computer runs.
As expected, the ULR detector is by far the most efficient. Among those that provide the estimation of the labels, it is seen that detector-B is uniformly the less demanding, and auction-sp is the most expensive.

\begin{table}[h]
\caption{Execution time normalized to ULR detector}
\begin{center}
\begin{tabular}{||c||c|c||c|c||}
\multicolumn{5}{c}{\textbf{auction-sp/ULR}} \\ 
\multicolumn{1}{c}{} &\multicolumn{2}{c||}{$\cH_0$}  & \multicolumn{2}{c}{$\cH_1$} \\
\cline{2-5} 
\multicolumn{1}{c||}{}& $n=10$ & $n=10^2$ &  $n=10$ & $n=10^2$ \\ \hline 
 $m=5$ &1666&13163&1486 &11607 \\ \hline
$m=20$ &15855&64099&14509 & 63426\\ \hline
\multicolumn{2}{c}{} \\ 
\multicolumn{5}{c}{\textbf{detector-A/ULR}} \\ 
\multicolumn{1}{c}{} &\multicolumn{2}{c||}{$\cH_0$}  & \multicolumn{2}{c}{$\cH_1$} \\
\cline{2-5} 
\multicolumn{1}{c||}{}&$n=10$ & $n=10^2$ &  $n=10$ & $n=10^2$ \\ \hline 
 $m=5$ & 65 & 1052& 67 & 1000 \\ \hline
$m=20$ &97&800&111 & 816 \\ \hline 
\multicolumn{2}{c}{} \\
\multicolumn{5}{c}{\textbf{detector-B/ULR}} \\ 
\multicolumn{1}{c}{} &\multicolumn{2}{c||}{$\cH_0$}  & \multicolumn{2}{c}{$\cH_1$} \\
\cline{2-5} 
\multicolumn{1}{c||}{}&$n=10$ & $n=10^2$ &  $n=10$ & $n=10^2$ \\ \hline 
 $m=5$ & 32 & 431 & 31 & 434 \\ \hline
$m=20$ &71&417&76 & 433 \\ \hline 
\end{tabular}
\end{center}
\label{tab:1}
\end{table}

\begin{figure}
\centering 
\includegraphics[width =220pt]{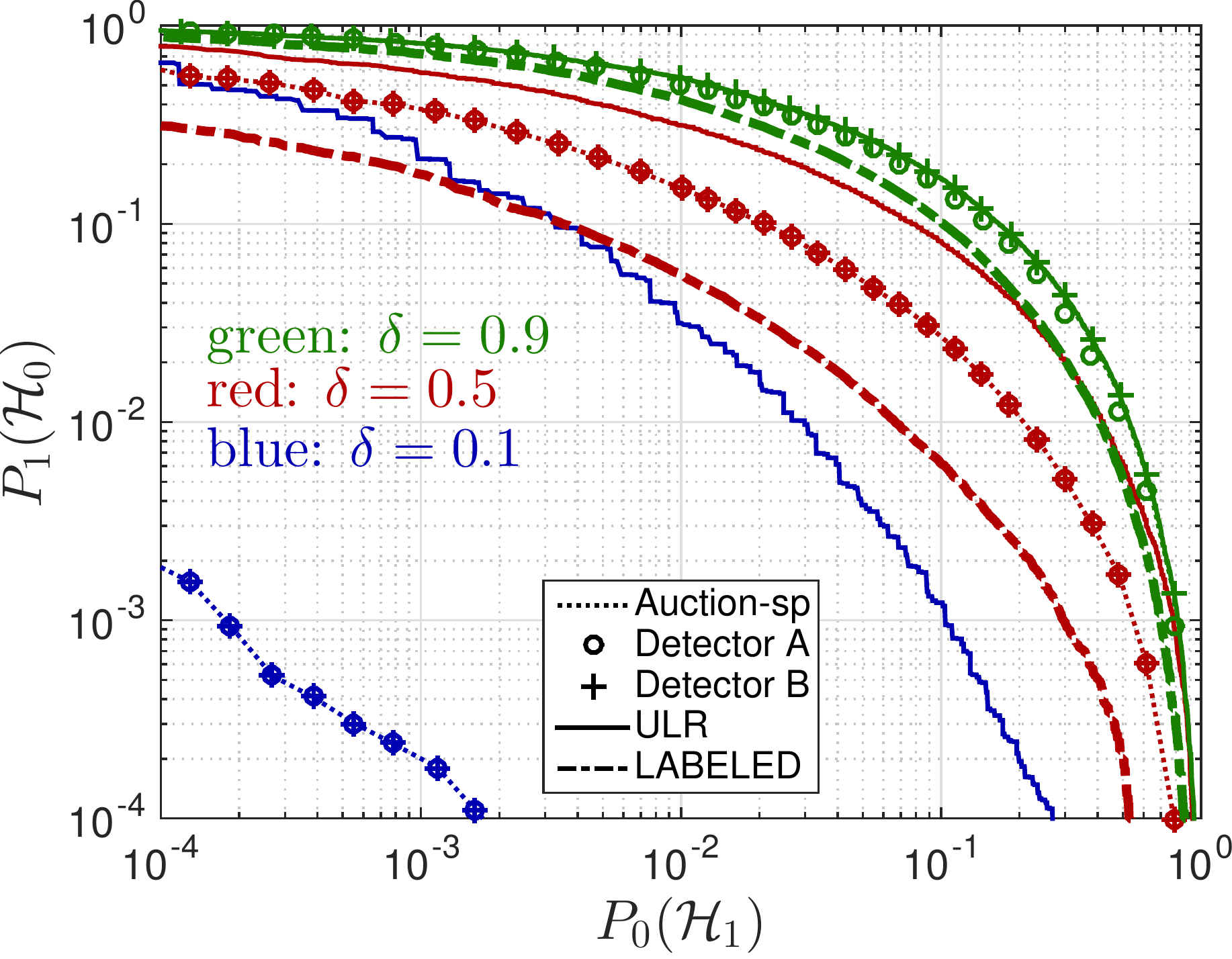}
 \caption{Second computer experiment. Type~II error probability versus type~I error probability for $m=5$, $n=20$, and three values of $\delta$.}
      \label{fig:3}
 \end{figure}

The relative ordering of the detectors' performance depends on the specific detection problem, and the performance assessment requires a case-by-case analysis. For instance, the superiority of the ULR detector shown in Figs.~\ref{fig:1} and~\ref{fig:2} is not a general rule, as shown in the second case study, which we now describe. Suppose that under $\cH_0$ data are \emph{iid} with common PMF $q_i=(\Delta, 2\Delta,\dots,m\Delta)^T$, where $\Delta=2/(m(m+1))$, and under $\cH_1$ the PMFs are as follows: for some $0<\delta<1$, the first $n/2$ distributions have mass $(1-\delta)$ at the first entry, and the remaining $n/2$ have mass $(1-\delta)$ at the last entry. In both cases all other entries have mass $\delta/(m-1)$. Figure~\ref{fig:3} reports the results of $10^5$ Monte Carlo simulations, for $m=5$, $n=20$, and $\delta=0.1, 0.5, 0.9$. When $\delta=0.9$ all the detectors perform similarly, but for smaller values of~$\delta$ the ULR detector is substantially outperformed by the other three detectors. For $\delta=0.1$ the ULR is extremely poor with respect to its competitors. For $\delta=0.5, 0.9$ the figure also shows, as benchmark, the ROC of the labeled detector. When performance improves, the loss incurred by the lack of the data labels increases and, for $\delta=0.1$, the curve of the labeled detector is out of the axis range.

\begin{figure}
\centering 
\includegraphics[width =220pt]{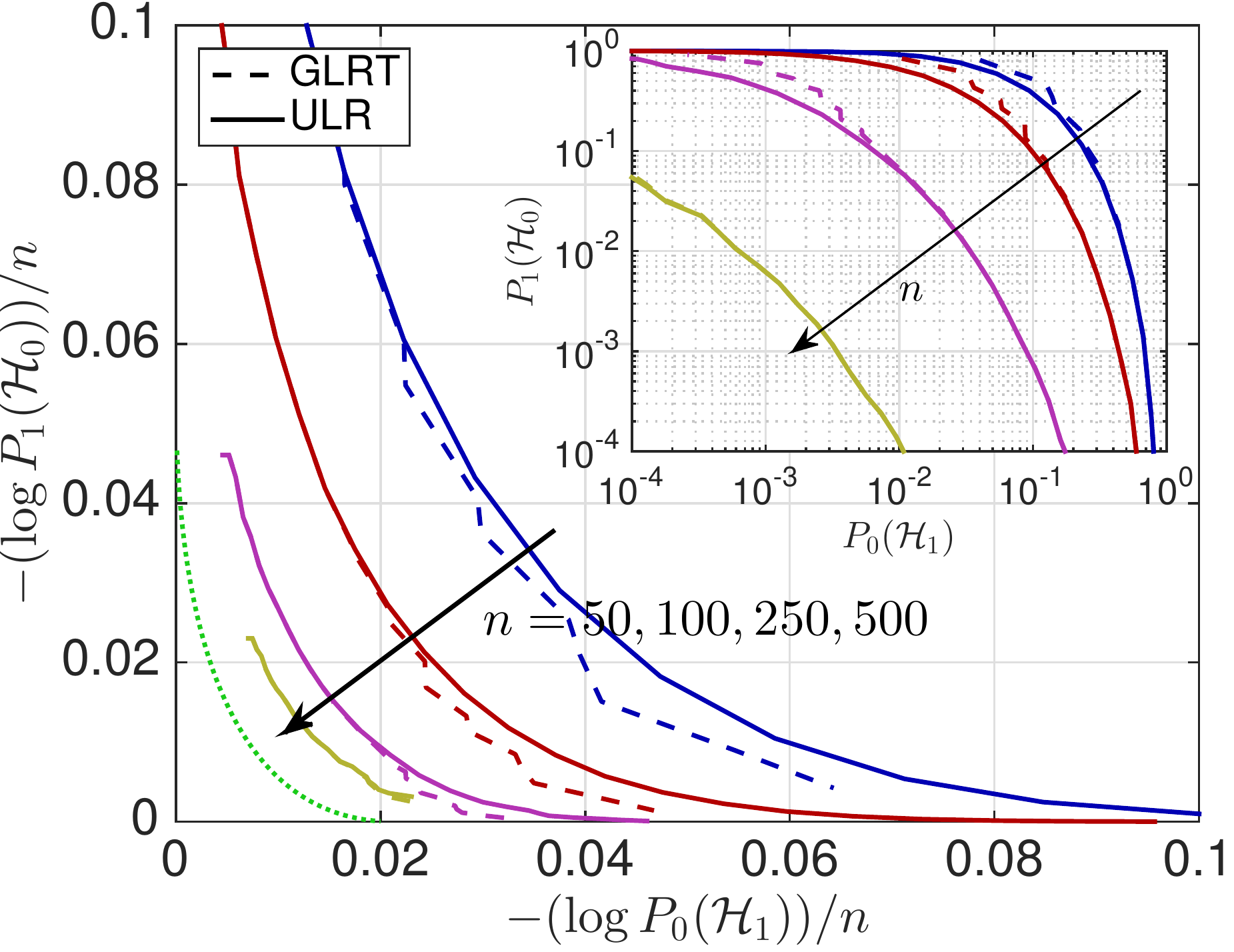}
 \caption{Error probabilities in the experiment with binary observations, $m=2$, and $n=50,100,250,500$. Also shown (green, dotted) is the curve $\Omega(\alpha)$ versus $\alpha$, see~(\ref{eq:csi}).}
      \label{fig:4}
 \end{figure}

Finally, let us consider the case of binary observations, $m=2$, and let us consider now non-identically distributed data under both hypotheses. Under $\cH_0$ we assume that the first $n/2$ data are drawn from distribution $(.5, .5)^T$, and the remaining $n/2$  from $(.3, .7)^T$.
Likewise, under $\cH_1$, the first half data come from the distribution $(.1, .9)^T$, and the other half from $(.9,.1)^T$.
The case of binary alphabets has some special features, which are investigated in~\cite{icassp2019submitted}. In particular, for binary alphabets, auction-sp, detector~A and detector~B, are exactly the same, and we accordingly report a single curve labeled as ``GLRT''.
The inset of Fig.~\ref{fig:4} shows that the ULR is worse than the GLRT for small $n$, but is essentially equivalent when $n$ grows. In the main plot, the same data of the inset are used but we depict $-\frac 1 n \log \P_1(\cH_0)$ versus $-\frac 1 n \log \P_0(\cH_1)$. Note that, in the main plot, for a fixed $n$, the ordering of the curves is reversed. 
On the same axes, for comparison, we also show (lowermost curve, dotted green) the curve $\Omega(\alpha)$, obtained by resolving numerically the convex optimization~(\ref{eq:csi}). The theoretical results of this article tell us that the optimal performance converges to $\Omega(\alpha)$.

Benchmark curves are not shown in Fig.~\ref{fig:4}, because the error probabilities of the labeled benchmark case are very small, and the number of Monte Carlo runs needed to estimating them reliably is much larger than that used in the figure.

\section{Conclusions}
\label{sec:conc}

We consider a canonical binary hypothesis test with independent data under both hypotheses.
Motivated by modern applications of sensor networks engaged in big data analysis, we assume that the observation vector $\bX^n=(X_1,\dots,X_n)$ collected by the peripheral units is 
delivered to the fusion center in the form of a random \emph{set} $\bX^n_u$, rather than a random vector. 
Namely, the values of the entries $\{X_i\}_{i=1}^n$ are known to the fusion center, but the positions that these values had in the original vector $\bX^n$ are not. 
The set $\bX^n_u$ is also known as the unlabeled version of $\bX^n$ and the problem is becoming known as \emph{detection by unlabeled data}.

The theoretical question addressed is how much information for detection is carried by $\bX^n_u$. We provide the asymptotic ($n\to\infty$) characterization of the performance of the optimal test in terms of an error exponent rate 
$\Omega(\alpha)$, which replaces the canonical rate $\Omega_{\rm lab}(\alpha)$ of the labeled case. It is proven that, when type~I error goes to zero as $\exp[-n \alpha]$ with the data size~$n$, type~II error may converge to zero as $\exp[-n \Omega(\alpha)]$ but not faster. The rate difference $\Omega_{\rm lab}(\alpha)-\Omega(\alpha)$ quantifies the loss of information induced by the loss of data labels (data positions). 

The second part of this paper addresses the practical question of how to solve the test by algorithms of affordable computational complexity and good performance. 
The ULR detector makes no attempts to estimate the labels and is very efficient computationally. If, aside from the decision, an estimate of the labels is also desired, we show that a viable detection algorithm for unlabeled data boils down to an assignment problem, for which a tailored form of the auction algorithm can be exploited. We also propose two alternative detection algorithms with good trade-off between performance and complexity, as we show by computer experiments.

For future studies it would be interesting to relate the performance and computational cost of the detection algorithms to the statistical distribution of the data, thus providing a theoretical assessment of their relative merits.
Our analysis is limited to the case in which the observations are independently drawn from a finite alphabet.
Generalization of the theoretical results to continuous random variables and designing practical algorithms for the continuous case are two important open problems.

\begin{appendices}
\numberwithin{equation}{section}

\section{Propositions~1 and 2: Sketch of proof}
\label{app:P1}

The assertion claimed in Proposition 1, when data are \emph{iid} under both hypotheses, can be found, e.g., in~\cite{csiszar-types}. 
A sketch of the proof in the case where data are not necessarily identically distributed exploits results from~\cite{blahut-HT&IT}, as follows.
Let
\beq
\Omega_{\rm lab}^{(n)}(\alpha) \dfz \inf_{\omega_{1:n}\in \cP(\cX^n): \, \frac 1 n \sum_{i=1}^n D(\omega_i \| q_i) \le \alpha}  \frac 1 n \sum_{i=1}^n D(\omega_i \| p_i),
\label{eq:Omegan}
\eeq
so that $\Omega_{\rm lab}(\alpha)=\lim_{n\rightarrow \infty} \Omega_{\rm lab}^{(n)}(\alpha)$, except for the fact that our definition 
of $\Omega_{\rm lab}(\alpha)$ in~(\ref{eq:Psi}) involves open divergence balls. 
Suppose that
$\liminf_{n\rightarrow \infty} - \frac 1 n \log \P_0(\bX^n \not \in A_n) \ge \alpha$. 
This implies that for all sufficiently small $\epsilon>0$ and all sufficiently large $n$, one has
$- \frac 1 n \log \P_0(\bX^n \not \in A_n) \ge \alpha - \epsilon$. For any $\gamma \in (0,1)$, $\alpha - \epsilon \ge \alpha - 2 \epsilon - \frac 1 n \log \gamma$, provided that
$n$ is large enough. Then,  Corollary~2 in~\cite{blahut-HT&IT} gives 
\begin{align}
- \frac 1 n \log \P_1(\bX^n \in A_n)  
  \; \;  \le - \frac{1}{n} \log \left (1-\frac{C(\alpha,\epsilon)}{n \epsilon^2}-\gamma \right)  + \Omega_{\rm lab}^{(n)}(\alpha-3 \epsilon)+\epsilon,
\end{align}%
where the constant $C(\alpha,\epsilon)\ge 0$ is made explicit in~\cite{blahut-HT&IT} and is of no concern to us.
Taking the limit superior this implies $\limsup_{n\rightarrow \infty} - \frac 1 n \log \P_1(\bX^n \in A_n) \le \Omega_{\rm lab}(\alpha)$, 
because $\epsilon$ is arbitrarily small, yielding~(\ref{eq:class1}).

Part $b)$ follows immediately by Corollary~1 in~\cite{blahut-HT&IT}, where it is shown that there exists a sequence $A_n^\ast$ of acceptance regions for $\cH_0$ such that
{\small \[
- \frac 1 n \log \P_0(\bX^n \not \in A_n^\ast) \ge \alpha, \quad 
- \frac 1 n \log \P_1(\bX^n \in A_n^\ast) \ge \Omega_{\rm lab}^{(n)}(\alpha).
\]}%
The former yields (\ref{eq:class3}) and, by  part a), also implies  
$\limsup_{n\rightarrow \infty}- \frac 1 n \log  \P_1(\bX^n \in A_n^\ast)$ $\le \Omega_{\rm lab}(\alpha)$. The latter gives \[\liminf_{n\rightarrow \infty} - \frac 1 n \log  \P_1(\bX^n \in A_n^\ast)  \ge \Omega_{\rm lab}(\alpha). \] Combining these two bounds gives~(\ref{eq:class4}). 

From the definition in~(\ref{eq:Omegan}) it is also seen that for $\Omega_{\rm lab}^{(n)}(0)= \frac 1 n \sum_{i=1}^n D(q_i \| p_i)$, and $\Omega_{\rm lab}^{(n)}(\alpha)=0$ for $\alpha\ge\frac 1 n \sum_{i=1}^n D(p_i \| q_i)$. Letting $n\to\infty$ these relationships give $\Omega_{\rm lab}(0)=\bar D(q_{1:\infty} \| p_{1:\infty})$, and $\Omega_{\rm lab}(\alpha)=0$ for $\alpha\ge \bar D(p_{1:\infty} \| q_{1:\infty})$.

Consider next proposition 2, which is a version of Chernoff-Stein's Lemma~\cite{CT2} applied to test~(\ref{eq:test}). The proof is based on defining, for a given integer $n$ and some $\epsilon>0$, the set of ``typical'' sequences $\bx^n$ that verify $\left | \frac 1 n \log \frac{q_{1:n}(\bx^n)}{p_{1:n}(\bx^n)} -\bar{D}(q_{1:\infty}\|p_{1:\infty}) \right | \le \epsilon$. Then, straightforward modifications of the arguments provided in~\cite[Chap.\ 11.8]{CT2} prove the claim.

\section{Proof of Theorem 1}
\label{app:proper}

\subsection{Properties of $\varphi_{\cH_h}(\lambda; r_i)$ and its Legendre transform}

To simplify the notation in the following we occasionally write $\varphi_{\cH_h}(\lambda)$ in place of $\varphi_{\cH_h}(\lambda; r_i)$, and similar.
Recall the definition $\cX^\prime = \cX \setminus\{x^\prime\}$, and the convention $\lambda(x^\prime)=0$. 
Under hypothesis $\cH_h$, $h=0,1$, let us consider the function $\varphi_{\cH_h}(\lambda)$ defined over $\Re^{|\cX|-1}$, see in~(\ref{eq:baq}).
The entries of gradient vector $\nabla \varphi_{\cH_h}(\lambda)$ are, for $x\in \cX^\prime$,
\begin{align}
\left [ \nabla \varphi_{\cH_h}(\lambda) \right ]_x = \frac{r_i(x)e^{\lambda(x)}}{\sum_{y\in \cX} r_i(y) e^{\lambda(y)}  } .
\label{eq:grado}
\end{align}
Let us regard the gradient vector as a mapping. Its domain is $\Re^{|\cX|-1}$ and its range is the convex set
\[\left \{ \omega \in \Re^{|\cX|-1}: \omega(x) > 0, \sum_{x \in \cX^\prime} \omega(x) < 1 \right \},\]
which is the projection of  ${\rm rin}(\cP(\cX))$ onto its $|\cX|-1$ coordinates $\omega(x)$, $x \in \cX^\prime$, where ${\rm rin}(A)$ denotes the relative interior of set~$A$~\cite{boyd-vandenberghe}. 
By adding the coordinate $\omega(x^\prime) = 1- \sum_{x \in \cX^\prime} \omega(x)$, the range of the mapping becomes ${\rm rin}(\cP(\cX))$, namely
$\nabla \varphi_{\cH_h} : \Re^{|\cX|-1} \mapsto {\rm rin}(\cP(\cX))$. Note also that, with this notational convention, 
$\varphi_{\cH_h}(0)=r_i \in {\rm rin}(\cP(\cX))$. 

The $(x,z)$-th entry of the Hessian matrix $\nabla^2 \varphi_{\cH_h}(\lambda)$ is
{\small \begin{align}
\left [ \nabla^2 \varphi_{\cH_h}(\lambda) \right ]_{x,z}= \left [\sum_{y\in \cX} r_i(y) e^{\lambda(y)}  \right ]^{-2}  
\left \{ 
\begin{array}{ll}
r_i(x)e^{\lambda(x)} \big [r_i(x^\prime)+\sum_{\begin{smallmatrix}y \in \cX^\prime \\ y\neq x \end{smallmatrix}} 
r_i(y) e^{\lambda(y)}  \big ], &  x = z, \\ 
-r_i(x)r_i(z)e^{\lambda(x)} e^{\lambda(z)}, & x \neq z.
\end{array}
\right . \label{eq:issess}
\end{align}}%
This shows that $\varphi_{\cH_h}(\lambda)$ is twice continuously differentiable throughout $\Re^{|\cX|-1}$. Straightforward algebra also shows that the Hessian matrix in~(\ref{eq:issess})
is strictly diagonally dominant, which implies that it is positive definite~\cite{Johnson-Horn}.
This proves that $\varphi_{\cH_h}(\lambda)$ is strictly convex over $\Re^{|\cX|-1}$~\cite{boyd-vandenberghe}.

A convex function is proper if it is $<\infty$ for at least one point and never takes the value $-\infty$; for a proper convex function, closedness is the same as lower semi-continuity~\cite{rockafellar-book}.
Thus, the convex function $\varphi_{\cH_h}(\lambda)$ is proper and closed because finite and everywhere continuous throughout $\Re^{|\cX|-1}$.

Recall that a proper convex function~$f$ is \emph{essentially smooth} if~\cite{rockafellar-book}:
$(i)$ $D_\lambda={\rm in}({\rm dom}(f))$ is nonempty, where ${\rm dom}(f)$ is the effective domain (the domain where $f$ is finite) and ${\rm in}(\cdot)$ denotes the interior;
$(ii)$ $f$ is differentiable throughout~$D_\lambda$; $(iii)$ $\lim_{k\to \infty} |\nabla f(\lambda_k)|=\infty$, whenever $\lambda_1,\lambda_2,\dots$ is a sequence in 
$D_\lambda$ converging to a boundary point of $D_\lambda$. A convex function on $\Re^{|\cX|-1}$ which is everywhere finite and differentiable throughout $\Re^{|\cX|-1}$ is essentially smooth.
Therefore, $\varphi_{\cH_h}(\lambda)$ is closed, proper, strictly convex, and essentially smooth, which allows us to invoke 
the following result, adapted from~\cite[Th. 26.5]{rockafellar-book}: 

\vspace*{3pt}
\noindent
\textsc{Theorem} (Facts from convex analysis):
\emph{Let $f(\lambda)$ be a closed proper convex function, and let 
\beq
F(\omega)= \sup_{\lambda \in \Re^{|\cX|-1}}\hspace*{-2pt} \left \{ \sum_{x\in \cX^\prime} \lambda(x) \omega(x) - f(\lambda)\right \}.
\label{eq:ltf}
\eeq
Let $D_\lambda={\rm in}({\rm dom}(f))$
and $D_\omega={\rm in}({\rm dom}(F))$. $F$ is strictly convex and essentially smooth on $D_\omega$ if and only if $f$ is strictly convex and essentially smooth on $D_\lambda$.
In this case: $(i)$ $(D_\omega,F)$ is the Legendre transformation of $(D_\lambda,f)$, and vice-versa; $(ii)$~the gradient mapping $\nabla f$ is continuous and one-to-one
from the open convex set $D_\lambda$ onto the open convex set $D_\omega$; $(iii)$~$\nabla F$ is the continuous inverse mapping of $\nabla f$:
$\nabla F=(\nabla f)^{-1}$. 
Function $F(\omega)$ in~(\ref{eq:ltf}) admits the representation:
$F(\omega)=\sum_{x\in \cX^\prime}  \omega(x) \nabla F(\omega) - f(\nabla F(\omega))$.} 

\vspace*{5pt}\noindent
Let $\Phi_{\cH_h}(\omega)$ be the Legendre transform of $\varphi_{\cH_h}(\lambda)$ defined by~(\ref{eq:ltf}).
We see that ${\rm dom}(\Phi_{\cH_h})=\cP(\cX)$ and $\Phi_{\cH_h}(\omega)$ is strictly convex and essentially smooth on ${\rm rin}(\cP(\cX))$. The mapping 
$\nabla \varphi_{\cH_h}: \Re^{|\cX|-1} \mapsto {\rm rin}(\cP(\cX))$ is one-to-one, with inverse $\nabla \Phi_{\cH_h}=(\nabla \varphi_{\cH_h})^{-1} : {\rm rin}(\cP(\cX)) \mapsto \Re^{|\cX|-1}$.

Assumption A in Sec.~\ref{sec:unlabeled} ensures that all these conclusions apply to the pair $(\psi_{\cH_h}(\lambda),
\Psi_{\cH_h}(\omega))$: Since $\psi_{\cH_h}(\lambda)$ is finite, strictly convex and twice continuously differentiable throughout $\Re^{|\cX|-1}$, we have that
${\rm dom}(\Psi_{\cH_h})=\cP(\cX)$, $\Psi_{\cH_h}(\omega)$ is strictly convex and essentially smooth on ${\rm rin}(\cP(\cX))$, and the mapping 
$\nabla \psi_{\cH_h}: \Re^{|\cX|-1} \mapsto {\rm rin}(\cP(\cX))$ is one-to-one, with inverse $\nabla \Psi_{\cH_h}=(\nabla \psi_{\cH_h})^{-1} : {\rm rin}(\cP(\cX)) \mapsto \Re^{|\cX|-1}$.

Assumption A also ensures $\nabla \psi_{\cH_h}(0)=\bar r$, which implies  $\nabla \Psi_{\cH_h}(\bar r)=0$. Since $\Psi_{\cH_h}(\omega)$
is strictly convex, this function has a unique global minimum at $\bar r \in {\rm rin}(\cP(\cX))$. By the representation of the Legendre transform given in the above
theorem, it is also easily seen that $\Psi_{\cH_h}(\bar r)=-\psi_{\cH_h}(0)=0$.

\vspace*{-12pt}
\subsection{Properties of $\Omega(\alpha)$}

Using the results of the previous section, the properties of the error exponent defined in~(\ref{eq:csi}) can be easily derived.  First, note that,
for $\alpha>0$:
{\small{
\begin{align}
\Omega(\alpha)= \hspace*{-10pt}\inf_{\omega \in \cP(\cX) \; : \; \Psi_{\cH_0}(\omega)<\alpha} \hspace*{-10pt} \Psi_{\cH_1}(\omega) 
=\hspace*{-10pt} \min_{\omega \in \cP(\cX) \; : \; \Psi_{\cH_0}(\omega)\le\alpha} \hspace*{-10pt}\Psi_{\cH_1}(\omega), \label{eq:may}
\end{align}}}%
where the second equality in~(\ref{eq:may}) follows by observing that $\Psi_{\cH_1}(\omega)$ is continuous and the set where the infimum is computed can be replaced by the 
compact set $\{\omega \in \cP(\cX)  :  \Psi_{\cH_0}(\omega)\le\alpha\}$, so that the minimum is attained at some point of this compact set~\cite{Zorichbook}.

Now, pick two positive values $\alpha_1$ and $\alpha_2$, and let $\omega_1,\omega_2 \in \cP(\cX)$ be the minimizers that attain $\Omega(\alpha_1)$ and $\Omega(\alpha_2)$, respectively.  
Let $\omega_\theta=\theta \omega_1+ (1-\theta) \omega_2$, with $0 \le \theta \le 1$. Clearly $\omega_\theta \in \cP(\cX)$
because $\cP(\cX)$ is convex. 
From the convexity of $\Psi_{\cH_0}(\omega)$ we have 
$\Psi_{\cH_0}(\omega_{\theta})$ $\le$ $\theta \Psi_{\cH_0}(\omega_1)+(1-\theta)\Psi_{\cH_0}(\omega_2)$ 
$\le$ $\theta \alpha_1+ (1-\theta) \alpha_2\dfz \alpha_\theta$. 
Thus, 
$\omega_{\theta} \in \{ \omega \in\cP(\cX) : \Psi_{\cH_0}(\omega)\le\alpha_\theta \}$. Then, from the convexity of $\Psi_{\cH_1}(\omega)$
\begin{align}
\Omega(\alpha_\theta)= \min_{\omega \in \cP(\cX) \; : \; \Psi_{\cH_0}(\omega)\le\alpha_\theta} \Psi_{\cH_1}(\omega) 
 \, \le \Psi_{\cH_1}(\omega_\theta) \le \theta \Psi_{\cH_1}(\omega_1)+(1-\theta) \Psi_{\cH_1}(\omega_2)  
 \, = \theta \Omega(\alpha_1) +(1-\theta) \Omega(\alpha_2),
\end{align}
which proves the convexity of $\Omega(\alpha)$.

It is immediate to see that $\Omega(\alpha)$ is nonincreasing in $\alpha$. We also see that 
$\alpha\ge\Psi_{\cH_0}(\bar p)$ $\Rightarrow$ $\Omega(\alpha)=0$, because in this case the set 
in~(\ref{eq:may}) where the minimum is computed includes $\bar p$, and $\Psi_{\cH_1}(\bar p)=0$.
At the origin, we define $\Omega(0)=\Psi_{\cH_1}(\bar q)$ by continuity.
Combining convexity and the nonincreasing property, we conclude that $\Omega(\alpha)$ is 
convex and strictly decreasing for $0< \alpha < \Psi_{\cH_0}(\bar p)$.

Next, by Jensen's inequality 
\beq
\lim_{n\to\infty}\frac 1 n \sum_{i=1}^n \log \sum_{x \in \cX} r_i(x) e^{\lambda(x)} \le \log \sum_{x \in \cX} \bar r(x) e^{\lambda(x)}, 
\eeq
which proves $\psi_{\cH_h}(\lambda; r_{1:\infty}) \le \psi_{\cH_h}(\lambda; \bar r)$, $\forall \lambda \in \Re^{|\cX|-1}$. 
For the Legendre transforms the inequality is reversed, yielding $\Psi_{\cH_h}(\omega; r_{1:\infty}) \ge \Psi_{\cH_h}(\omega; \bar r)$, $\forall \omega \in \cP(\cX)$.
Noting that $\Psi_{\cH_h}(\omega; \bar r)=\Phi_{\cH_h}(\omega; \bar r)$, 
we then get, 
for $\alpha > 0$,
\begin{align}
\Omega(\alpha; p_{1:\infty},q_{1:\infty})&= \inf_{\omega \in \cP(\cX) \; : \; \Psi_{\cH_0}(\omega; q_{1:\infty})<\alpha} \Psi_{\cH_1}(\omega; p_{1:\infty})
\nonumber \\
&  \ge \inf_{\omega \in \cP(\cX) \; : \; \Psi_{\cH_0}(\omega; q_{1:\infty})<\alpha} \Psi_{\cH_1}(\omega; \bar p) \quad  
\left [ = \Omega(\alpha; \bar p,q_{1:\infty})\right ]\nonumber \\
&  \ge \inf_{\omega \in \cP(\cX) \; : \; \Psi_{\cH_0}(\omega; \bar q)<\alpha} \Psi_{\cH_1}(\omega; \bar p) \qquad \;  \left [ = \Omega(\alpha; \bar p, \bar q)\right ], 
\nonumber 
\end{align}
and similarly one ontains $\Omega(\alpha; p_{1:\infty},q_{1:\infty})\ge \Omega(\alpha; p_{1:\infty}, \bar q)$.
All the inequalities of~(\ref{eq:mi}) are so proved, except the first, which follows by the operational meaning of the two rate error functions $\Omega(\alpha)$
and $\Omega_{\rm lab}(\alpha)$ provided by Proposition~1 and Theorem~2.

Finally, we show that $\Psi_{\cH_h}(\omega; \bar r)=D(\omega || \bar r)$, $h=0,1$.
For any $\omega \in \cP(\cX)$, any $\lambda \in \Re^{|\cX|-1}$, and $\lambda(x^\prime)=0$: 
\begin{align}
\psi_{\cH_h}(\lambda; \bar r) &= \log \sum_{x \in \cX} \bar r(x) e^{\lambda(x)} =\log \sum_{x \in \cX} \omega(x) \frac{\bar r(x)e^{\lambda(x)}}{\omega(x)} \nonumber \\
& \ge \sum_{x \in \cX} \omega(x) \log \frac{\bar r(x)e^{\lambda(x)} }{\omega(x)}
= \sum_{x \in \cX^\prime}  \omega(x) \lambda(x)  - D(\omega \| \bar r), \nonumber  
\end{align}
yielding $\sum_{x \in \cX^\prime} \omega(x) \lambda(x) - \psi_{\cH_h}(\lambda; \bar r) \le D(\omega \| \bar r)$. 
If there exists a vector $\lambda \in \Re^{|\cX|-1}$ such that this upper bound is achieved, then $\Psi_{\cH_h}(\omega; \bar r)=D(\omega \| \bar r)$
because of the definition of Legendre transform, see~(\ref{eq:ltf}). Direct substitution shows that such vector is
$\lambda(x)= \log\frac{\omega(x) \, \bar r(x^\prime)}{\bar r(x) \, \omega(x^\prime)}$, $x \in \cX^\prime$.

\section{Proof of Theorem~2}
\label{app:P2}

The type vector $t_{\bX^n}(x)$, $x \in \cX$, defined in~(\ref{eq:tv}) contains only $|\cX|-1$ independent components. 
Here we work with the \emph{reduced} type vector $t_{\bX^n}^\prime$ obtained by deleting the entry $t_{\bX^n}(x^\prime)$ from $t_{\bX^n}$.
Accordingly, let us introduce the set $\cQ(\cX^\prime)=\{\omega(x), x\in \cX^\prime : \omega\in\cP(\cX) \}$ of probability vectors $\omega \in \cP(\cX)$ from which the entry $\omega(x^\prime)$ is deleted. For notational simplicity we loosely use the same symbol $\omega$ to denote vectors in $\cP(\cX)$, vectors in $\cQ(\cX^\prime)$, and vectors in $\Re^{|\cX|-1}$. Also, as done in Appendix~\ref{app:proper}, we occasionally omit to make explicit the dependence 
of the various functions on the underlying statistical distributions.

There exists an obvious one-to-one correspondence between $(|\cX|-1)$-vectors in $\cQ(\cX^\prime)$ and $|\cX|$-vectors in $\cP(\cX)$, as well as between reduced type $t_{\bX^n}^\prime$ and type $t_{\bX^n}$. Thus, the event $\{ t_{\bX^n}^\prime \in E^\prime \}$ is the same of $\{ t_{\bX^n} \in E \}$, provided that $E\in \cP(\cX)$ is the element that corresponds to $E^\prime\in\cQ(\cX^\prime)$.

Let $\cH_h$, $h=0,1$, be the hypothesis in force, and recall that the distribution of $X_i$ under $\cH_h$ is denoted by $r_i \in \cP(\cX)$, with $i=1,2,\dots$.
Let $\lambda \in \Re^{|\cX|-1}$, and consider the logarithmic moment generating function of the reduced type vector $t_{\bX^n}^\prime$:
\begin{align}
\Lambda_{\cH_h,n}(\lambda)&\dfz 
\log \E_h \exp \left\{ \sum_{x\in \cX^\prime} \lambda(x) t_{\bX^n}^\prime(x) \right\} 
= \sum_{i=1}^n \log  \sum_{x\in \cX} r_i(x) e^{\frac{\lambda(x)}{n}}, \label{eq:bGEass}
\end{align}
where, we recall,  $\lambda(x^\prime)=0$ by convention. Note that $\lim_{n\to\infty} \frac 1 n  \Lambda_{\cH_h,n}(n\lambda)$ is exactly the function $\psi_{\cH_h}(\lambda)$
defined  in~(\ref{eq:newpsi}).
Its Legendre transform $\Psi_{\cH_h}(\omega)$ is defined in~(\ref{eq:legt}) for $\omega \in \cP(\cX)$. We use the same symbol
$\Psi_{\cH_h}(\omega)$ to denote the Legendre transform of $\psi_{\cH_h}(\lambda)$ as function of the $(|\cX|-1)$-vector 
$\omega \in \cQ(\cX^\prime)$, in which case the definition is extended to all $\Re^{|\cX|-1}$ by setting $\Psi_{\cH_h}(\omega)=\infty$ for 
$\omega \in \Re^{|\cX|-1} \setminus \cQ(\cX^\prime)$.

The proof of Theorem 2 is based on the following version of G\"artner-Ellis theorem, see~\cite[Th.\ 3.2.6]{DZbook} or~\cite[Th.\ V.6]{DenHollander}. 

\vspace*{3pt}
\noindent
\textsc{Theorem} (G\"artner-Ellis)
\emph{Suppose that the function $\psi_{\cH_h}(\lambda)$ in~(\ref{eq:newpsi}) is finite and differentiable throughout~$\Re^{|\cX|-1}$. Then for any set 
$A^\prime \subseteq \Re^{|\cX|-1}$
we have the large deviation principle:
\begin{align}
\inf_{\omega \in {\rm cl}(A^\prime)} \Psi_{\cH_h}(\omega) \le \liminf_{n\to\infty} - \frac 1 n \log \P_h(t_{\bX^n}^\prime \in A^\prime) 
\le \limsup_{n\to\infty} - \frac 1 n \log \P_h(t_{\bX^n}^\prime \in A^\prime) \le \inf_{\omega \in {\rm in}(A^\prime)} \Psi_{\cH_h}(\omega),
\label{eq:LDPGE}
\end{align}}%

\noindent
In~(\ref{eq:LDPGE}) and in what follows ${\rm in}(A^\prime)$ and ${\rm cl}(A^\prime)$ denote the interior and the closure of $A^\prime$, respectively. 
The complement of $A^\prime$ will be denoted by $\overline{A^\prime}$.
These operations are relative to~$\Re^{|\cX|-1}$.

By Assumption A, $\psi_{\cH_h}(\lambda)$ in~(\ref{eq:newpsi}) is finite and differentiable in $\Re^{|\cX|-1}$, which allows us to apply the G\"artner-Ellis theorem.
Let $E \subseteq \cP(\cX)$ be an arbitrary closed acceptance region for $\cH_0$ and let $E^\prime$ be the corresponding closed set in $\cQ(\cX^\prime)$.
Note that $\overline{E^\prime}$ is an open set $\in \Re^{|\cX|-1}$.
Under $\cH_0$:
\begin{subequations}
\begin{align}
  \liminf_{n\to\infty} - \frac 1 n \log \P_0(t_{\bX^n} \in \overline{E}) 
&= \liminf_{n\to\infty} - \frac 1 n \log \P_0(t_{\bX^n}^\prime \in \overline{E^\prime}) \nonumber \\
&   \le \limsup_{n\to\infty} - \frac 1 n \log \P_0(t_{\bX^n}^\prime \in \overline{E^\prime})  \label{eq:ch-1} \\
&  \le \inf_{\omega\in   \overline{E^\prime} \subseteq \Re^{|\cX|-1}} \Psi_{\cH_0}(\omega)  \label{eq:ch0in}, \\
&  \le \Psi_{\cH_0}(\omega), \quad \forall \omega \in \overline{E^\prime} \subseteq \Re^{|\cX|-1} \label{eq:ch0in2},
\end{align} \label{eq:ch0}%
\end{subequations}%
where:~(\ref{eq:ch0in}) follows by the upper bound in~(\ref{eq:LDPGE}) for the open set $\overline{E^\prime}$.

For $\alpha>0$, let us impose $\liminf_{n\to\infty} - \frac 1 n \log \P_0(t_{\bX^n} \in \overline{E}) \ge \alpha$. 
From~(\ref{eq:ch0in2}) this implies
$\Psi_{\cH_0}(\omega) \ge \alpha$, $\forall \omega \in \overline{E^\prime} \subseteq \Re^{|\cX|-1}$,
so that $\Psi_{\cH_0}(\omega) <\alpha \Rightarrow \omega \in E^\prime \subseteq \cQ(\cX^\prime)$, and therefore 
\begin{align}
C_\alpha^\prime \dfz \{ \omega \in \cQ(\cX^\prime) : \Psi_{\cH_0}(\omega)<\alpha \} \subseteq E^\prime.
\label{eq:setin2}
\end{align} 
Under $\cH_1$, we have
\begin{subequations}
\begin{align}
\limsup_{n\to\infty} - \frac 1 n \log \P_1(t_{\bX^n} \in E)  &=\limsup_{n\to\infty} - \frac 1 n \log \P_1(t_{\bX^n}^\prime \in E^\prime) &  \nonumber \\
& \le \limsup_{n\to\infty} - \frac 1 n \log \P_1(t_{\bX^n}^\prime \in C_\alpha^\prime) & \label{eq:bobo}\\
& \le \inf_{\omega \in {\rm in}(C_\alpha^\prime) \subseteq \cQ(\cX^\prime)} \Psi_{\cH_1}(\omega) &\label{eq:bobo2} \\
& = \inf_{\omega \in C_\alpha^\prime \subseteq \cQ(\cX^\prime)} \Psi_{\cH_1}(\omega) 
=\inf_{\omega \in C_\alpha \subseteq \cP(\cX)} \Psi_{\cH_1}(\omega) &\label{eq:bobo4} 
\end{align}
\end{subequations}%
where~(\ref{eq:bobo}) follows by $C_\alpha^\prime\subseteq E^\prime$,~(\ref{eq:bobo2}) follows by the upper bound in~(\ref{eq:LDPGE}), 
and the first equality in~(\ref{eq:bobo4}) is obtained by the continuity of $\Psi_{\cH_1}(\omega)$ on $\cQ(\cX^\prime)$.
This proves part~a).

To prove part b), let us set $E^{\prime\ast}={\rm cl}(C_\alpha^\prime)= \{ \omega \in \cQ(\cX^\prime) : \Psi_{\cH_0}(\omega)\le\alpha \}$,
and let $E^\ast$ be its corresponding set in $\cP(\cX)$.
Under $\cH_0$: 
\begin{subequations}
\begin{align}
\liminf_{n\to\infty} - \frac 1 n \log \P_0(t_{\bX^n} \in \overline{E^\ast})  
&=\liminf_{n\to\infty} - \frac 1 n \log \P_0(t_{\bX^n}^\prime \in \overline{E^{\prime\ast}}) \nonumber \\
&   \ge  \hspace*{-7pt} \inf_{\omega \in \Re^{|\cX|-1} : \Psi_{\cH_0}(\omega)\ge\alpha}  \hspace*{-12pt} \Psi_{\cH_0}(\omega) 
 =  \hspace*{-5pt} \inf_{\omega \in \cQ(\cX^\prime) : \Psi_{\cH_0}(\omega)\ge\alpha}  \hspace*{-10pt} \Psi_{\cH_0}(\omega) \label{eq:flowb3} \\
&=  \hspace*{-5pt} \inf_{\omega \in \cP(\cX) : \Psi_{\cH_0}(\omega)\ge\alpha}  \hspace*{-10pt}\Psi_{\cH_0}(\omega) \ge \alpha, \label{eq:flowb5}
\end{align}%
\end{subequations}%
where the inequality in~(\ref{eq:flowb3}) is the lower bound in~(\ref{eq:LDPGE}), and the equality in~(\ref{eq:flowb3}) can be verified by considering separately 
the two cases $\alpha$ such that
$\{\omega \in \cQ(\cX^\prime) : \Psi_{\cH_0}(\omega)\ge\alpha\} = \emptyset$ (the infimum over the empty set being $\infty$ by definition),
and $\neq \emptyset$.
This proves~(\ref{eq:t3a}).
Finally, under $\cH_1$, note that
\begin{align}
\hspace*{-12pt} \inf_{\omega \in {\rm in}(E^{\prime\ast})} \Psi_{\cH_1}(\omega)= \hspace*{-5pt}
\inf_{\omega \in {\rm cl}({\rm in}(E^{\prime\ast}))} \hspace*{-10pt} \Psi_{\cH_1}(\omega)= \hspace*{-5pt}
\inf_{\omega \in {\rm cl}(E^{\prime\ast})} \hspace*{-10pt} \Psi_{\cH_h}(\omega) \label{eq:sun}
\end{align}
where the first equality follows by the continuity of $\Psi_{\cH_1}(\omega)$ on $\cQ(\cX^\prime)$ 
and the second follows by ${\rm cl}(E^{\prime\ast}) = {\rm cl}({\rm in}(E^{\prime\ast}))$.
From~(\ref{eq:sun}) we see that the lower and the upper bounds in~(\ref{eq:LDPGE}) coincide, and the large deviation principle gives a precise limit:
\begin{subequations}\begin{align}
\lim_{n\to\infty} - \frac 1 n \log \P_1(t_{\bX^n} \in \ast) &=\lim_{n\to\infty} - \frac 1 n \log \P_1(t_{\bX^n}^\prime \in E^{\prime\ast}) \nonumber \\
&=  \hspace*{-5pt}  \inf_{\omega \in \cQ(\cX^\prime) : \Psi_{\cH_0}(\omega)\le\alpha}  \hspace*{-10pt}  \Psi_{\cH_1}(\omega) \label{eq:tiok2} \\
&=  \hspace*{-5pt}  \inf_{\omega \in \cP(\cX) : \Psi_{\cH_0}(\omega)\le\alpha}  \hspace*{-10pt}  \Psi_{\cH_1}(\omega)
=  \hspace*{-5pt}  \inf_{\omega \in \cP(\cX) : \Psi_{\cH_0}(\omega)<\alpha}  \hspace*{-10pt}  \Psi_{\cH_1}(\omega), \label{eq:tiok4}
\end{align}\end{subequations}
where~(\ref{eq:tiok2}) follows by~(\ref{eq:LDPGE}) and~(\ref{eq:sun}), while the second equality in~(\ref{eq:tiok4}) follows by the continuity of $\Psi_{\cH_1}(\omega)$ on $\cP(\cX)$.


\section{Asymptotics of $t_{\bX^n}$}
\label{app:wp1}

Let $\widetilde \bX^n=(\widetilde X_1,\dots,\widetilde X_n)$, with the entries $\widetilde X_i$ drawn \emph{iid} from $\bar r$, and let $t_{\widetilde \bX^n}$ be the corresponding type.
For $t_{\widetilde \bX^n}$ the standard strong law of large numbers~\cite[Th. 22.1]{billingsley-book2} gives, $\forall x \in \cX$, $t_{\widetilde \bX^n}(x) \to \bar r(x)$
with probability one. As to $t_{\bX^n}$ --- the type when data are drawn from $r_{1:n}$ --- note that ${\rm VAR}_h[\cI(X_i=x)]=r_i(x)(1-r_i(x))$, where ${\rm VAR}_h$ denotes the variance computed under $\cH_h$, and 
$
\sum_{i=1}^\infty {\rm VAR}_h[\cI(X_i=x)]/i^2 \le \sum_{i=1}^\infty i^{-2}/4 = \pi^2/24 < \infty.
$
This implies the following convergence with probability one~\cite[Th.\ 1.14]{shao}:
\beq
\frac 1 n \sum_{i=1}^n \cI(X_i=x) -  \frac 1 n \sum_{i=1}^n \E_h [\cI(X_i=x)] \rightarrow  0,
\eeq
from which we see that $t_{\bX^n}(x) \to \bar r(x)$ exactly as does $t_{\widetilde \bX^n}(x)$.
By triangular inequality, for any $\epsilon>0$, and all sufficiently large $n$, $|t_{\bX^n}(x)-t_{\widetilde \bX^n}(x)|<\epsilon$ with probability~one.

\end{appendices}



\begin{thebibliography}{10}
\providecommand{\url}[1]{#1}
\csname url@samestyle\endcsname
\providecommand{\newblock}{\relax}
\providecommand{\bibinfo}[2]{#2}
\providecommand{\BIBentrySTDinterwordspacing}{\spaceskip=0pt\relax}
\providecommand{\BIBentryALTinterwordstretchfactor}{4}
\providecommand{\BIBentryALTinterwordspacing}{\spaceskip=\fontdimen2\font plus
\BIBentryALTinterwordstretchfactor\fontdimen3\font minus
  \fontdimen4\font\relax}
\providecommand{\BIBforeignlanguage}[2]{{%
\expandafter\ifx\csname l@#1\endcsname\relax
\typeout{** WARNING: IEEEtran.bst: No hyphenation pattern has been}%
\typeout{** loaded for the language `#1'. Using the pattern for}%
\typeout{** the default language instead.}%
\else
\language=\csname l@#1\endcsname
\fi
#2}}
\providecommand{\BIBdecl}{\relax}
\BIBdecl

\bibitem{ICASSP2018unlab}
S.~Marano and P.~Willett, ``Sometimes they come back: Testing two simple
  hypotheses (in the realm of unlabeled data),'' in \emph{Proc.\ of the 2018
  IEEE International Conference on Acoustics, Speech and Signal Processing
  ({ICASSP} 2018)}, Calgary, Alberta, Canada, April 15-20 2018.

\bibitem{mahler_book}
R.~Mahler, \emph{Statistical Multisource-Multitarget Information Fusion}.\hskip
  1em plus 0.5em minus 0.4em\relax Artech House, 2007.

\bibitem{mahler_tutorial}
------, ``Statistics 101 for multisensor, multitarget data fusions,''
  \emph{{IEEE} Transactions on Aerospace and Electronic Systems}, vol.~19,
  no.~1, pp. 53--64, Jan. 2004.

\bibitem{Humphreys-2008}
T.~E. Humphreys, B.~M. Ledvina, M.~L. Psiaki, B.~W. O'Hanlon, and P.~M.
  {Kintner, Jr.}, ``Assessing the spoofing threat: Development of a portable
  {GPS} civilian spoofer,'' in \emph{2016 IEEE Conference on Communications and
  Network Security (CNS)}, Savanna, GA, Sep. 16-19 2008, pp. 2314--2325.

\bibitem{6415619-Milcom2012}
Q.~Zeng, H.~Li, and L.~Qian, ``{GPS} spoofing attack on time synchronization in
  wireless networks and detection scheme design,'' in \emph{MILCOM 2012 - 2012
  IEEE Military Communications Conference}, Oct 2012, pp. 1--5.

\bibitem{7860525CNS2016}
P.~Pradhan, K.~Nagananda, P.~Venkitasubramaniam, S.~Kishore, and R.~S. Blum,
  ``{GPS} spoofing attack characterization and detection in smart grids,'' in
  \emph{2016 IEEE Conference on Communications and Network Security (CNS)}, Oct
  2016, pp. 391--395.

\bibitem{6400273-SMARTGRID2013}
Z.~Zhang, S.~Gong, A.~D. Dimitrovski, and H.~Li, ``Time synchronization attack
  in smart grid: Impact and analysis,'' \emph{IEEE Transactions on Smart Grid},
  vol.~4, no.~1, pp. 87--98, March 2013.

\bibitem{Challa2003}
S.~Challa, R.~J. Evans, and X.~Wang, ``A {Bayesian} solution and its
  approximations to out-of-sequence measurement problems,'' \emph{Information
  Fusion}, no.~4, pp. 185--199, 2003.

\bibitem{4608943-AC2008}
L.~Schenato, ``Optimal estimation in networked control systems subject to
  random delay and packet drop,'' \emph{IEEE Transactions on Automatic
  Control}, vol.~53, no.~5, pp. 1311--1317, June 2008.

\bibitem{7266584-BracaWillett2015}
L.~M. Millefiori, P.~Braca, K.~Bryan, and P.~Willett, ``Adaptive filtering of
  imprecisely time-stamped measurements with application to {AIS} networks,''
  in \emph{2015 18th International Conference on Information Fusion (Fusion)},
  July 2015, pp. 359--365.

\bibitem{Vetterli2018IT}
J.~Unnikrishnan, S.~Haghighatshoar, and M.~Vetterli, ``Unlabeled sensing with
  random linear measurements,'' \emph{{IEEE} Transactions on Information
  Theory}, vol.~64, no.~5, pp. 3237--3253, May 2018.

\bibitem{7447086-ALLERTON2015}
------, ``Unlabeled sensing: Solving a linear system with unordered
  measurements,'' in \emph{2015 53rd Annual Allerton Conference on
  Communication, Control, and Computing (Allerton)}, Sept 2015, pp. 786--793.

\bibitem{eldar}
Y.~C. Eldar, \emph{Sampling Theory, Beyond Bandlimited Systems}, 2nd~ed.\hskip
  1em plus 0.5em minus 0.4em\relax Cambridge, U.K.: Cambridge University Press,
  2015.

\bibitem{6853755-ICASSP2014}
V.~Emiya, A.~Bonnefoy, L.~Daudet, and R.~Gribonval, ``Compressed sensing with
  unknown sensor permutation,'' in \emph{2014 IEEE International Conference on
  Acoustics, Speech and Signal Processing (ICASSP)}, May 2014, pp. 1040--1044.

\bibitem{SLAM}
S.~Thrun and J.~J. Leonard, ``Simultaneous localization and mapping,'' in
  \emph{Springer Handbook of Robotics}, B.~Siciliano and O.~Khatib, Eds.\hskip
  1em plus 0.5em minus 0.4em\relax Heidelberg: Springer Berlin, 2008, pp.
  871--889.

\bibitem{8234626-Caire2018SP}
S.~Haghighatshoar and G.~Caire, ``Signal recovery from unlabeled samples,''
  \emph{IEEE Transactions on Signal Processing}, vol.~66, no.~5, pp.
  1242--1257, March 2018.

\bibitem{Abid2017}
\BIBentryALTinterwordspacing
A.~Abid, A.~Poon, and J.~Zou. (2017, May 4) Linear regression with shuffled
  labels. [Online]. Available: \url{http://arxiv.org/abs/1705.01342}
\BIBentrySTDinterwordspacing

\bibitem{Pananjady2017}
\BIBentryALTinterwordspacing
A.~Pananjady, M.~J. Wainwright, and T.~A. Courtade. (2017, April 24) Denoising
  linear models with permutated data. [Online]. Available:
  \url{http://arxiv.org/abs/1704.07461}
\BIBentrySTDinterwordspacing

\bibitem{7852261-ALLERTON2016}
------, ``Linear regression with an unknown permutation: Statistical and
  computational limits,'' in \emph{2016 54th Annual Allerton Conference on
  Communication, Control, and Computing (Allerton)}, Sept 2016, pp. 417--424.

\bibitem{Pananjady2016}
\BIBentryALTinterwordspacing
------. (2016, August 9) Linear regression with an unknown permutation:
  statistical and computational limits. [Online]. Available:
  \url{https://arxiv.org/abs/1608.02902}
\BIBentrySTDinterwordspacing

\bibitem{WangSP2018}
G.~Wang, J.~Zhu, R.~S. Blum, P.~Willett, S.~Marano, V.~Matta, and P.~Braca,
  ``Signal amplitude estimation and detection from unlabeled binary quantized
  samples,'' \emph{{IEEE} Transactions on Signal Processing}, vol.~66, no.~16,
  pp. 4291--4303, Aug. 2018.

\bibitem{7904659-CommLett2017}
J.~Zhu, H.~Cao, C.~Song, and Z.~Xu, ``Parameter estimation via unlabeled
  sensing using distributed sensors,'' \emph{IEEE Communications Letters},
  vol.~21, no.~10, pp. 2130--2133, Oct 2017.

\bibitem{marano-ICASSP17}
S.~Marano, V.~Matta, P.~Willett, P.~Braca, and R.~Blum, ``Hypothesis testing in
  the presence of {Maxwell's} daemon: Signal detection by unlabeled
  observations,'' in \emph{Proc.\ of the IEEE International Conference on
  Acoustics, Speech and Signal Processing ({ICASSP} 2017)}, New Orleans, LA,
  USA, 5-9 Mar. 2017.

\bibitem{keller-2009}
L.~Keller, M.~J. Siavoshani, C.~Fragouli, K.~Argyraki, and S.~Diggavi,
  ``Identity aware sensor networks,'' in \emph{Proc.\ of the 26th IEEE
  International Conference on Computer Communications ({INFOCOM} 2009)}, Rio De
  Janeiro, Brazil, April, 19-25 2009, pp. 2177--2185.

\bibitem{ludo}
Y.~M. Lu and M.~N. Do, ``A theory for sampling signals from a union of
  subspaces,'' \emph{{IEEE} Transactions on Signal Processing}, vol.~56, no.~6,
  pp. 2334--2345, Jun. 2008.

\bibitem{CT2}
T.~M. Cover and J.~A. Thomas, \emph{Elements of Information Theory},
  2nd~ed.\hskip 1em plus 0.5em minus 0.4em\relax New Jersey, USA:
  Wiley-Interscience, 2006.

\bibitem{blahut-HT&IT}
R.~E. Blahut, ``Hypothesis testing and information theory,'' \emph{{IEEE}
  Transactions on Information Theory}, vol.~20, no.~4, pp. 405--417, Jul. 1974.

\bibitem{Lehmann-testing3}
E.~L. Lehmann and J.~P. Romano, \emph{Testing Statistical Hypotheses},
  3rd~ed.\hskip 1em plus 0.5em minus 0.4em\relax Springer, 2005.

\bibitem{boyd-vandenberghe}
S.~Boyd and L.~Vandenberghe, \emph{Convex Optimization}.\hskip 1em plus 0.5em
  minus 0.4em\relax Cambridge, UK: Cambridge University Press, 2004.

\bibitem{rockafellar-book}
R.~T. Rockafellar, \emph{Convex analysis}.\hskip 1em plus 0.5em minus
  0.4em\relax Princeton, NJ: Princeton University Press, 1970.

\bibitem{kaydetection}
S.~M. Kay, \emph{Fundamentals of Statistical Signal Processing, Volume II:
  Detection Theory}.\hskip 1em plus 0.5em minus 0.4em\relax Englewood Cliffs,
  New Jersey: Prentice Hall, 1998.

\bibitem{bertsekas-tsitsiklis}
D.~P. Bertsekas and J.~N. Tsitsiklis, \emph{Parallel and Distributed
  Computation: Numerical Methods}.\hskip 1em plus 0.5em minus 0.4em\relax
  Belmont, MA: Athena Scientific, 1997.

\bibitem{Kuhn-Hungarian}
H.~W. Kuhn, ``The {Hungarian} method for the assignment problem,'' \emph{Nav.
  Res. Log. Q.}, vol.~2, pp. 83--97, 1955.

\bibitem{Munkres}
J.~Munkres, ``Algorithms for the assignment and transportation problems,''
  \emph{Journal of the Society for Industrial and Applied Mathematics}, vol.
  5(1), pp. 32--38, 1957.

\bibitem{blackman-popoli}
S.~Blackman and R.~Popoli, \emph{Design and Analysis of Modern Tracking
  Systems}.\hskip 1em plus 0.5em minus 0.4em\relax Artech House, 1999.

\bibitem{bertsekas-netopt}
D.~P. Bertsekas, \emph{Network Optimization: Continuous and Discrete
  Models}.\hskip 1em plus 0.5em minus 0.4em\relax Belmont, MA: Athena
  Scientific, 1998.

\bibitem{bertsekas-castanon}
D.~P. Bertsekas and D.~A. Castanon, ``The auction algorithm for the
  transportation problem,'' \emph{Annals of Operations Research}, vol.~20, pp.
  67--96, 1989.

\bibitem{JVC}
R.~Jonker and A.~Volgenant, ``Improving the {Hungarian} assignment algorithm,''
  \emph{Operations Research Letters}, vol.~5, pp. 171--175, 1986.

\bibitem{icassp2019submitted}
S.~Marano and P.~Willett, ``Making decisions with shuffled bits,'' in
  \emph{Proc.\ of the 2019 IEEE International Conference on Acoustics, Speech
  and Signal Processing ({ICASSP} 2019)}, Brighton, UK, 12--17 May,
  \emph{submitted.}

\bibitem{csiszar-types}
I.~Csisz\'ar, ``The method of types,'' \emph{{IEEE} Transactions on Information
  Theory}, vol.~44, no.~6, pp. 2505--2523, Oct. 1998.

\bibitem{Johnson-Horn}
R.~Horn and C.~Johnson, \emph{Matrix Analysis}.\hskip 1em plus 0.5em minus
  0.4em\relax Cambridge, UK: Cambridge University Press, 1985.

\bibitem{Zorichbook}
V.~A. Zorich, \emph{Mathematical Analysis {I}}.\hskip 1em plus 0.5em minus
  0.4em\relax Berlin: Springer, 2004.

\bibitem{DZbook}
A.~Dembo and O.~Zeitouni, \emph{Large Deviations Techniques and Applications},
  2nd~ed.\hskip 1em plus 0.5em minus 0.4em\relax New York: Springer, 1998.

\bibitem{DenHollander}
F.~den Hollander, \emph{Large Deviations}, ser. Fields Institute
  Monographs.\hskip 1em plus 0.5em minus 0.4em\relax Providence, Rhode Island:
  American Mathematical Society, 2000.

\bibitem{billingsley-book2}
P.~Billingsley, \emph{Probability and Measure}, 3rd~ed.\hskip 1em plus 0.5em
  minus 0.4em\relax New York: Wiley-Interscience, 1995.

\bibitem{shao}
H.~Shao, \emph{Mathematical Statistics}, 2nd~ed.\hskip 1em plus 0.5em minus
  0.4em\relax Springer, 2003.

\end{thebibliography}
\end{document}